\documentclass[journal]{IEEEtran}
\IEEEoverridecommandlockouts
\usepackage{cite}
\usepackage{amsmath,amssymb,amsfonts}
\usepackage{graphicx}
\usepackage{textcomp}
\usepackage[colorlinks,linkcolor=red,anchorcolor=blue,citecolor=blue]{hyperref}
\usepackage{url}
\usepackage{booktabs}
\usepackage{nicefrac}
\usepackage{microtype}
\usepackage[table,xcdraw]{xcolor}
\usepackage{multirow}
\usepackage{makecell}
\usepackage{enumitem}
\usepackage{caption}
\usepackage{pifont}
\usepackage{threeparttable}
\usepackage{subfigure}
\usepackage{subcaption}
\usepackage[T1]{fontenc}
\usepackage{courier}
\usepackage{algorithm}
\usepackage{algpseudocodex}

\definecolor{header_color}{RGB}{78, 121, 167}
\definecolor{row_color_2}{RGB}{237, 237, 237}
\DeclareMathAlphabet\mathbfcal{OMS}{cmsy}{b}{n}
\newcommand{\ten}[1]{\mathbfcal{#1}}
\newcommand{\mat}[1]{\mathbf{#1}}

\begin{document}

\title{
  FETTA: Flexible and Efficient Hardware Accelerator for Tensorized Neural Network Training
  \thanks{This work is co-funded by Intel Strategic Research Sectors (SRS) - Systems Integration SRS \& Devices SRS. (\textit{Corresponding author: Zheng Zhang.})}
}

\author{ Jinming Lu, Jiayi Tian, Hai Li, \\ Ian Young, \textit{Fellow, IEEE}, Zheng Zhang
\thanks{
J. Lu, J. Tian, and Z. Zhang are with Department of Electrical and Computer Engineering, University of California, Santa Barbara, CA 93106; H. Li and I. Young are with Intel Corporation, Hillsboro, OR 97124; (email: jinminglu@ucsb.edu; zhengzhang@ece.ucsb.edu).
}
}

\maketitle

\begin{abstract}

    The increasing demand for on-device training of deep neural networks (DNNs) aims to leverage personal data for high-performance applications while addressing privacy concerns and reducing communication latency. However, resource-constrained platforms face significant challenges due to the intensive computational and memory demands of DNN training.
    Tensor decomposition emerges as a promising approach to compress model size without sacrificing accuracy. Nevertheless, training tensorized neural networks (TNNs) incurs non-trivial overhead and severe performance degradation on conventional accelerators due to complex tensor shaping requirements.
    To address these challenges, we propose FETTA, an algorithm and hardware co-optimization framework for efficient TNN training. On the algorithm side, we develop a contraction sequence search engine (CSSE) to identify the optimal contraction sequence with the minimal computational overhead. On the hardware side, FETTA features a flexible and efficient architecture equipped with a reconfigurable contraction engine (CE) array to support diverse dataflows.
    Furthermore, butterfly-based distribution and reduction networks are implemented to perform flexible tensor shaping operations during computation.
    Evaluation results demonstrate that FETTA achieves reductions of $20.5\times$/$100.9\times$, $567.5\times$/$45.03\times$, and $11609.7\times$/$4544.8\times$ in terms of processing latency, energy, and energy-delay product (EDP) over GPU and TPU, respectively.
    Moreover, working on the tensorized training, FETTA outperforms prior accelerators with a speedup of $3.87\sim 14.63\times$, and an energy efficiency improvement of $1.41 \sim 2.73\times$ on average.
\end{abstract}


  \maketitle

  \section{Introduction}

    Deep neural networks (DNNs) have achieved remarkable success in various real-world applications, particularly in computer vision and natural language processing. Traditionally, DNNs are trained on high-performance graphic processing units (GPUs) and subsequently deployed on personal devices for real-world use.
    To optimize models for deployment, techniques such as quantization \cite{nagel2021white}, pruning \cite{blalock2020state}, and tensor decomposition \cite{liu2023tensor} are widely employed to reduce model size and computational demands.
    However, there has been a growing need for DNNs to continuously learn from new data after deployment. This capability is essential to mitigate performance degradation caused by distribution distortion between training and deployment datasets while safeguarding user privacy by eliminating the need to transfer data to cloud servers~\cite{kwontinytrain,zhu2023pockengine, lin2022device}. Consequently,  the design of efficient on-device training architectures have become a critical focus of research~\cite{lin2023tiny, zhu2024device, kim2024dacapo, yu20248}.

    However, on-device training presents significantly greater challenges compared to inference~\cite{lee2021overview, khouas2024training}.
    Training involves higher computational complexity (at least $3\times$ more computation), increased memory demands, and more diverse computational patterns.
    Existing solutions for accelerating DNN training on resource-constrained devices often employ reconfigurable engines or sparsity techniques to address these challenges~\cite{Zhao2016FCNNAF, Liu2017AnFP,yang2020procrustes, wang2022swpu, lu2022theta}.  Although some methods effectively reduce computational complexity, they often introduce accuracy degradation, complicated sparsity indexing mechanisms, or suboptimal utilization of hardware resources.

    The tensor decomposition algorithm has demonstrated significant promise as a model compression technique, achieving high compression ratios while maintaining accuracy~\cite{liu2023tensor, wang2023tensor, novikov2015tensorizing, ye2020block}.
    A network compressed using tensor decomposition is referred to as a tensorized neural network (TNN), which consists of a combination of uncompressed layers and tensorized layers.
    {
    In this paper, we refer to uncompressed fully-connected or convolutional layers as dense layers for clarity.
    }
    Recent studies have proven the practicality of training TNNs from scratch \cite{yang2024comera, ghiasvand2024communication, feng2024long, chekalina2023efficient, qiu2024compute, zangrandogeometry}, revealing their potential for acceleration during the training phase.
    While several hardware accelerators have been developed for TNN inference acceleration~\cite{gong2023ette, deng2019tie, gong2022algorithm,zhang2024tetrix}, applying tensor decomposition to accelerate the training process introduces new challenges.
    These challenges hinder performance and remain insufficiently explored.
    \begin{itemize}[leftmargin=*]
      \item[\ding{202}] Although tensorized layers significantly reduce parameters compared to dense layers, this does not directly translate into computational efficiency. A tensorized layer consists of a sequence of tensor contraction operations,
      potentially increasing computational efforts if executed in a straightforward computing scheme \cite{deng2019tie}.
      Moreover, tensorized layers produce additional intermediate results that must be stored for back-propagation, which diminishes memory reduction benefits if not handled properly.

      \item[\ding{203}] The training of TNNs requires high-order tensor contraction operations, with varying tensor orders across layers. In contrast, the  standard linear layer only operate on the fixed tensor order of 2. Consequently, tensor contraction requires complex tensor shaping to align the data layout and dataflow. Support for training and diverse computing schemes further complicates this requirements.
      As a result, mapping TNN training to existing accelerators often leads to severely degraded computational utilization and efficiency.
    \end{itemize}

    To address the above issues, this work proposes a flexible and efficient accelerator for TNN training through hardware and algorithm co-optimization. Our contributions are summarized as follows.
    \begin{itemize}[leftmargin=*]
      \item We propose a light-weight DNN training solution based on tensor decomposition. We analyze the impact of computing schemes for tensorized layers and highlight the inefficiencies of previous approaches.
      We develop a contraction sequence search engine (CSSE) to identify the optimal contraction sequence for achieving the best hardware performance.
      CSSE is built upon an enlarged search space and  employs a two-stage search strategy, taking into account both the optimality of search results and the time budget of search.
      \item We introduce a novel hardware architecture, FETTA, for TNN training. FETTA is a flexible and efficient hardware architecture to fully leverage the algorithmic benefits.
      FETTA features a hierarchical Contraction Engine (CE) array implemented on a transposable systolic array, enabling reconfigurable support for diverse dataflows in TNN training.
      To further enhance efficiency, butterfly-based distribution and reduction networks are implemented to facilitate flexible data shaping operations, eliminating the need for additional external memory access and avoiding bank conflicts.
      \item We implement FETTA under ASAP 7nm technology and evaluate it on multiple benchmarks. FETTA achieves a speedup of $20.5\times$ /$100.9\times$ and an energy efficiency improvement of $576.5\times$/$45.0\times$ over GPU and TPU on dense training workloads, respectively.
      Compared to state-of-the-art training accelerators on tensorized training workloads, FETTA still achieves $3.87\sim 14.63\times$ and $1.41 \sim 2.73\times$  improvements in terms of processing speed and energy efficiency, respectively.
    \end{itemize}.

    The remainder of this paper is organized as follows.
    Section \ref{sec:back} provides a brief introduction to tensor decomposition.
    Section \ref{sec:motivation} highlights the inefficiency of existing solutions and motivates our approach.
    Section \ref{sec:algorithm} presents the proposed contraction sequence search engine.
    Section \ref{sec:architecture} elaborates the design of the FETTA hardware architecture.
    Section \ref{sec:methods} describes the evaluation methodology, and Section \ref{sec:results} presents the evaluation results and compares FETTA with prior works.
    Finally, we draw the conclusion in Section \ref{sec:conclusion}.

    \section{Background}\label{sec:back}
    \begin{figure}[tbp]
      \centering
      \includegraphics[width=0.95\linewidth]{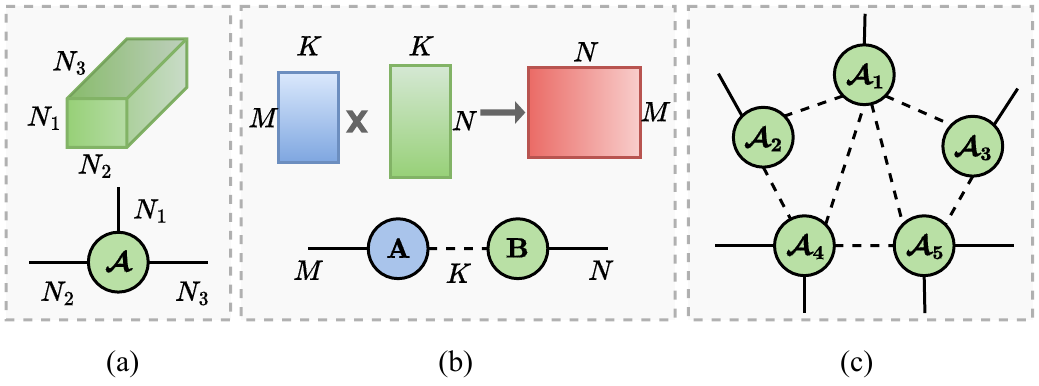}
      \vspace{-8pt}
      \caption{Tensor network diagrams illustrating (a) a 3rd-order tensor node, (b) a tensor contraction of the matrix multiplication, (c) a multi-node tensor network.}
      \label{fig:tensor}

      \vspace{-10pt}
    \end{figure}

    \begin{figure*}
      \centering
        \subfigure[]{
          \includegraphics[height=0.18\linewidth]{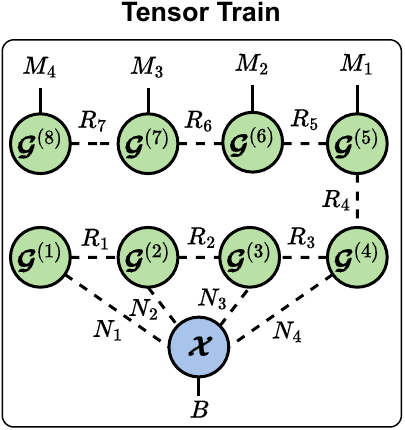}
          \label{subfig:tt_graph}
        }
        \subfigure[]{
          \includegraphics[height=0.18\linewidth]{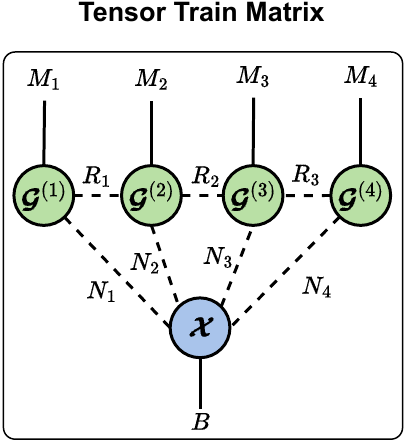}
          \label{subfig:ttm_graph}
        }
        \subfigure[]{
          \includegraphics[height=0.18\linewidth]{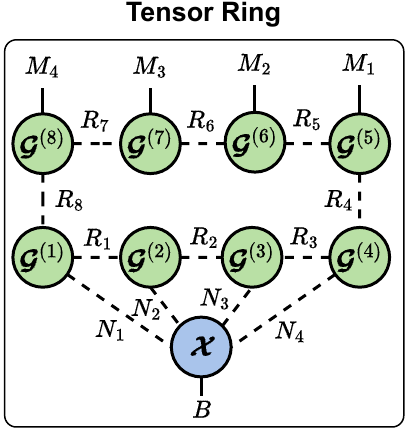}
          \label{subfig:tr_graph}
        }
        \subfigure[]{
          \includegraphics[height=0.18\linewidth]{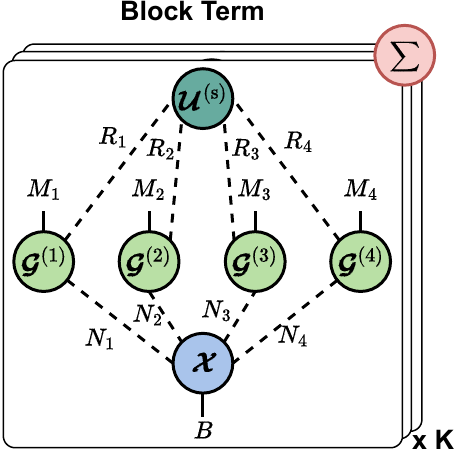}
          \label{subfig:bt_graph}
        }
        \subfigure[]{
          \includegraphics[height=0.18\linewidth]{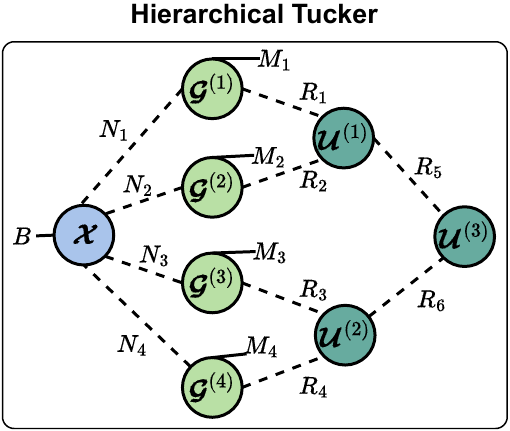}
          \label{subfig:ht_graph}
        }
        \vspace{-10pt}
        \caption{Tensor network diagrams for (a) Tensor Train, (b) Tensor Train Matrix, (c) Tensor Ring, (d) Block Term, (e) Hierarchical Tucker. In each graph, $\ten{X} \in \mathbb{R}^{B \times N_1 \times N_2 \times N_3 \times N_4}$, $\ten{W} \in \mathbb{R}^{M_1 \times M_2 \times M_3 \times M_4 \times N_1 \times N_2 \times N_3 \times N_4}$, and  $\ten{Y} \in \mathbb{R}^{B \times M_1 \times M_2 \times M_3 \times M_4}$}
      \label{fig:tensor_graph}

      \vspace{-10pt}
      \end{figure*}

    \subsection{Tensor Basis}
    \textbf{Tensor} is the most common terminology used to represent data in the current deep learning literature \cite{landsberg2011tensors, wang2023tensor}.
    A tensor is a multi-dimensional data array that generalizes vectors (1st-order tensors) and matrices (2nd-order tensors) to higher dimensions, where the order refers to the number of tensor dimensions.
    Throughout the paper, lower-case letters (\textit{e.g.}, a) denote scalars; lower-case bold letters (\textit{e.g.}, $\mat{a}$) denote vectors; upper-case bold letters (\textit{e.g.}, $\mat{A}$) denote matrices; upper-case calligraphic bold letters (\textit{e.g.}, $\ten{A}$) denote tensors (order $\geq$ 3).
    A $d$-th-order tensor with dimensions $N_1, N_2, \dots, N_d$ is denoted as $\ten{A} \in \mathbb{R}^{N_1 \times N_2 \dots \times N_d}$.

    \textbf{Tensor Networks} are structured collections of tensors interconnected through tensor contraction operations. The tensor network diagram \cite{penrose1971applications} is a graphical representation used to describe both the data and the operations within tensor networks.
    In this representation, a $d$-th-order tensor $\ten{A}$ is depicted as a node with$d$ edges, where the value associated with each edge represents the corresponding dimension size. Fig. \ref{fig:tensor}{(a)} illustrates the diagram of a 3rd-order tensor.

    \textbf{Tensor Contraction} occurs when two or more tensors with shared dimensions are merged (contracted) into a single tensor.
    During contraction, the connected edges between the tensors disappear, while the dangling edges remain.
    Considering a pair of tensors $\ten{A} \in \mathbb{R}^{M_1 \times ...\times M_t \times K_1 \times ... \times K_s}$ and $\ten{B} \in \mathbb{R}^{K_1 \times ... \times K_s \times N_1 \times ...\times N_d}$, the tensor contraction of $\ten{A}$ and $\ten{B}$ is formulated as Eq. (\ref{eq:contract_2}).
    \begin{equation}\label{eq:contract_2}
      \ten{C}_{[m_1..m_t, n_1..n_d]} =  \sum_{k_1..k_s } \ten{A}_{[m_1.. m_t, k_1..k_s] }  \ten{B}_{[k_1..k_s, n_1.. n_d]}.
    \end{equation}

    where $\ten{C} \in \mathbb{R}^{M_1\times M_2\times ...\times M_t \times N_1 \times N_2...\times N_d}$.

    The contraction of a matrix-matrix multiplication is illustrated in Fig. \ref{fig:tensor}{(b)} as an example.
    In addition, a sequence of tensor contractions among multiple tensors in a tensor network is shown in Fig. \ref{fig:tensor}{(c)}.
    \subsection{Tensor Decomposition}

    Tensor decomposition, equivalent to tensor networks in certain contexts, is a promising method for DNN compression~\cite{novikov2015tensorizing, ye2020block, liu2023tensor, hrinchuk2019tensorized}.
    A DNN compressed by tensor decomposition is called a tensorized neural network (TNN).
    In DNNs, linear layers generally dominate the number of parameters and computational overhead.
    A linear layer is typically formulated as $\mat{Y} = \mat{X} \mat{W}^T $, where $\mat{Y} \in \mathbb{R}^{B\times N}, \mat{W} \in \mathbb{R}^{M\times N}$, and $\mat{X} \in \mathbb{R}^{B\times M}$.

    In a TNN, a linear layer is first represented into the tensorized format, where all data matrices are reshaped into higher-order tensors. Specifically, $\mat{X}$ is tensorized into $\ten{X} \in \mathbb{R}^{B\times N_1 \times N_2 ...\times N_{t}}$, $\mat{Y}$ is denoted as $\ten{Y} \in \mathbb{R}^{B\times M_1 \times M_2 ...\times M_{s}}$, and $\mat{W}$ is denoted as $\ten{W} \in \mathbb{R}^{M_1\times M_2...\times M_{s} \times N_1 \times ...\times N_{t}}$, where $M = \prod_{i=1}^{s} M_i$, $N = \prod_{i=1}^{t} N_i$.
    Consequently, the computation of a tensorized layer is formulated by a tensor contraction as in Eq. (\ref{eq:tensor_layer}).
    \begin{equation}\label{eq:tensor_layer}
      \ten{Y}_{[b, m_1, ... m_{s}]} = \sum_{n_1...n_{t}} \ten{X}_{[b, n_1,... n_{t}]} \ten{W}_{[m_1, ... m_{s}, n_1,... n_{t}]}.
    \end{equation}
    The weight tensor is then decomposed into a set of small-scale core tensors $\{\ten{G}^{(i)}\}_{i=1}^d$, leading to a significant compression ratio while maintaining accuracy.

    There exist many tensor decomposition methods varying in topology, order, number, and representation ability. Here, we provide a brief overview of the commonly used tensor decomposition methods.

    \textbf{Tensor-Train (TT)} \cite{gong2023ette, yang2024comera} decomposes $\ten{W}$ into $d$ 3rd-order small-scale core tensors $\{\ten{G}^{(i)}\}_{i=1}^d$, where $\ten{G}^{(i)} \in \mathbb{R}^{R_{i-1}\times M_i \times R_i}$ when $ i \leq s$, $\ten{G}^{(i)} \in \mathbb{R}^{R_{i-1}\times N_{i-d} \times R_i}$ when $ i > s $, and $d = t + s$.
    Here $\{R_i\}_{i=0}^d$ are ranks of the core tensors, and $R_0 = R_d = 1$ by default.
    The tensor network diagram of the obtained TT layer is illustrated in Fig. \ref{fig:tensor_graph}.
    The computation is formulated as Eq. \eqref{eq:tt}.
    \begin{equation}\label{eq:tt}
      \begin{aligned}
      &\ten{W}_{[m_1, ... m_{s}, n_1, ... n_t]} = \\
      & \sum_{r_1...r_d} \ten{G}^{(1)}_{[r_0, m_1, r_1]}... \ten{G}^{(s)}_{[r_{s-1}, m_{s}, r_{s}]}
       \ten{G}^{(s+1)}_{[r_{s}, n_{1}, r_{s+1}]} ...\ten{G}^{(s+t)}_{[r_{d-1}, n_{t}, r_d]}.
      \end{aligned}
    \end{equation}

    \textbf{Tensor-Train Matrix (TTM)} is a variant of TT that is widely used in DNN compression \cite{novikov2015tensorizing, yang2017tensor, deng2019tie}, which decomposes a $2d$-th-order weight tensor $\ten{W} \in R^{M_1\times M_2 ...M_{s} \times N_1 \times ... \times N_{t}}$ into $d$ 4th-order core tensors $\ten{G}^{(i)} \in \mathbb{R}^{R_{i-1}\times M_i \times N_i \times R_i}$, where $d = s = t $, and $R_0=R_d=1$. The tensor network is shown in Fig. \ref{fig:tensor_graph}.
    \begin{equation}\label{eq:ttm}
      \begin{aligned}
         & \ten{W}_{[m_1,...m_d, n_1, ... n_d]} =                                                                                               \\
         & \sum_{r_1...r_d} \ten{G}^{(1)}_{[r_0,m_1, n_1, r_1]} \ten{G}^{(2)}_{[r_1,m_2, n_2, r_2]} ...\ten{G}^{(d)}_{[r_{d-1},m_d, n_d, r_d]}. \\
      \end{aligned}
    \end{equation}

    \textbf{Tensor Ring (TR)} \cite{zhao2016tensor, pan2019compressing} is another variant of TT, which links the endpoints of TT to construct a ring structure and brings a higher representation ability than TT. In addition, TR defines $R_0=R_d$ and makes them changeable.
    Fig. \ref{fig:tensor_graph} shows its graph and Eq. \eqref{eq:tr} describes its computation.
    \begin{equation}\label{eq:tr}
      \begin{aligned}
      &\ten{W}_{[m_1, ... m_{s}, n_1, ... n_{s}]} = \\
      & \sum_{r_1...r_d} \ten{G}^{(1)}_{[r_d, m_1, r_1]}... \ten{G}^{(s)}_{[r_{s-1}, m_{s}, r_{s}]}
       \ten{G}^{(s+1)}_{[r_{s}, n_{1}, r_{s+1}]} ...\ten{G}^{(s+t)}_{[r_{d-1}, n_{t}, r_d]}.
      \end{aligned}
    \end{equation}

    \textbf{Hierarchical Tucker (HT)} \cite{yin2020compressing} has a tree-like structure, which recursively decomposes the weight tensor $\ten{W}$ into $d$ 3rd-order core tensors $\ten{G}^{(i)} \in \mathbb{R}^{\times M_i \times N_i \times R_i}$ as leaf nodes and $k$ transfer tensors $\{\ten{U}^{(j)}\}_{j=1}^k$ as non-leaf nodes.
    As shown in Fig. \ref{fig:tensor_graph}, core tensors are responsible for computing with the input tensor, and transfer tensors only involve internal contraction operations.

    \textbf{Block Term (BT)} \cite{ye2018learning} realizes a trade-off between the canonical polyadic (CP) decomposition \cite{kolda2009tensor} and the Tucker decomposition \cite{de2008decompositions}. BT decomposes $\ten{W}$  into $K$ block terms, and each term conducts contraction between a $d$-th-order transfer tensor $\ten{U}^{(k)} \in \mathbb{R}^{R_1\times R_2...R_d}$ and $d$ 3rd-order core tensors $\ten{G}^{(k,i)} \in \mathbb{R}^{M_i \times N_i \times R_i}$, where $k \in [1, K]$ and $i \in [1, d]$. The tensor network diagram for BT is shown in Fig. \ref{fig:tensor_graph}. All block terms are summed to reconstruct the original weight tensor.

    \subsection{Training of DNNs}
    DNN training generally contains three main compute-intensive phases, including forward propagation (FP), backward propagation (BP), and weight gradient (WG)\cite{lu2022eta}.

    \textbf{Forward Propagation (FP)}: In the $i$-th layer, the output $Y_i$ is computed using the input $X_i$ and the weight $W_i$. The output $Y_i$ then serves as the input for the next layer.

    \textbf{Backward Propagation (BP)}: The input gradient $d\mat{X}_i$ is calculated by multiplying the output gradient $d\mat{Y}_i$ with the weight $W_i$. This gradient, $d\mat{X}_i$, is then propagated backward to the previous layer.

    \textbf{Weight Gradient (WG)}: The weight gradient $d\mat{W}_i$ is computed using the output gradient $d\mat{Y}_i$ and the input $X_i$. This gradient is subsequently used to update the weight $W_i$.

    The computations for these three phases in a linear layer can be expressed as shown in Eq. (\ref{eq:fc_train}).
    \begin{equation}\label{eq:fc_train}
      \begin{aligned}
        \mat{Y}_i           & = \mat{X}_i \mat{W}^T_i,           \\
        \textrm{d}\mat{X}_i & = \textrm{d}\mat{Y}_i \mat{W}_i ,  \\
        \textrm{d}\mat{W}_i & = \mat{X}^T_i \textrm{d}\mat{Y}_i.
      \end{aligned}
    \end{equation}

    \begin{figure}[bp]
      \centering
      \vspace{-10pt}
      \includegraphics[width=0.9\linewidth]{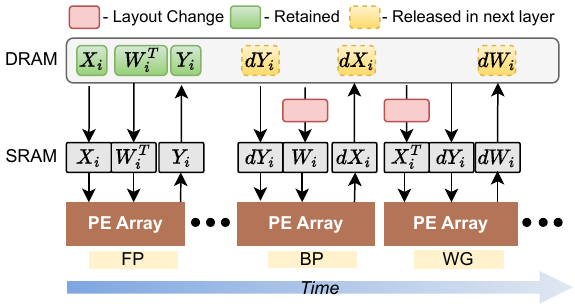}
      \caption{Processing flow diagram for DNN training.}

      \label{fig:train_flow}
    \end{figure}

    During training, the overall processing flow of each layer on a typical systolic array-based accelerator is depicted in Fig. \ref{fig:train_flow}.
    There are two important observations:
    \ding{202} The input activation $X_i$  generated during the FP phase must be retained in DRAM for an extended duration until the corresponding WG phase for that layer is completed. In tensorized layers, the intermediate results produced by tensor contraction steps must also be stored for WG computations, introducing additional overhead compared to standard layers.
    \ding{203} Although the same data is reused across in different phases, variations in computational patterns during different phases necessitate additional matrix transpose operations. For higher-order tensor formats, more complex data shaping operations, such as permutation and reshaping, are required.
    To handle this situation, either an on-chip or off-chip data layout reordering mechanism is essential.

    \section{Motivations}\label{sec:motivation}

    Tensor decomposition significantly reduces model parameters, offering substantial potential for accelerating DNN training and inference.
    However, there are still several challenges to fully translate its compression benefits into the improvements of hardware performance and efficiency.

    \subsection{Computing Schemes}

    When implementing and deploying models on hardware platforms, the number of parameters alone does not directly determine hardware performance.
    Even with an ideal hardware accelerator operating at full utilization, the computational demands are primarily determined by the number of floating-point operations (FLOPs) and the amount of data access required.
    Especially for training, because of the inherent characteristics of TNN, various computing schemes can be employed, and additional storage is often needed to accommodate intermediate results. Without proper management, hardware performance may even worse than dense training.

    A tensorized layer represented by a tensor network diagram comprises $K$ nodes, including an input node and $K-1$ weight nodes, requiring $K-1$ tensor contraction operations.
    Since the order in which these operations are performed does not affect the values of the final result, there are many possible execution sequences for a tensorized layer,
    These sequences can differ significantly in terms of FLOPs and memory access, leading to substantial variations in hardware efficiency.

\begin{figure}[t]
      \centering
      \includegraphics[width=\linewidth]{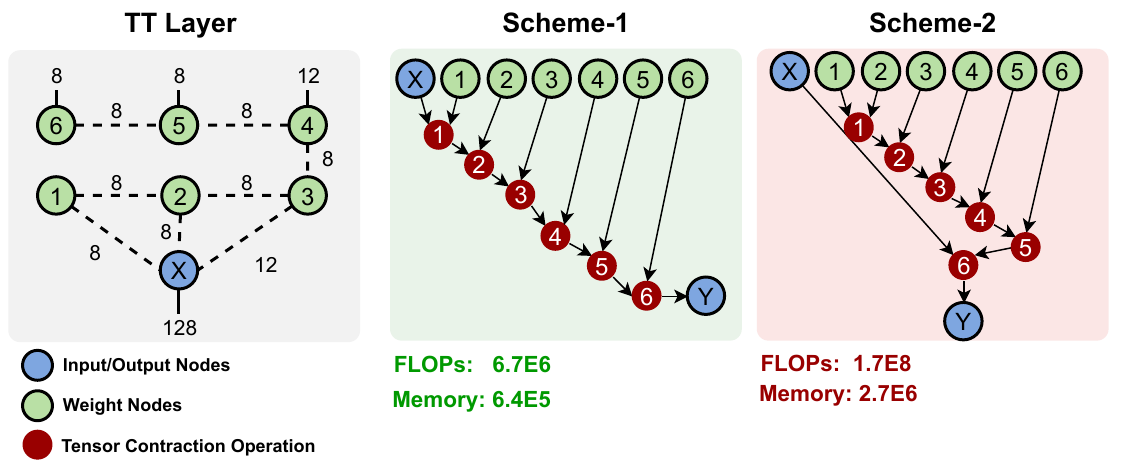}
      \caption{Example computing schemes for a TT layer. Weight nodes are denoted with index for simplicity.}
      \label{fig:prior_path}
      \vspace{-10pt}

    \end{figure}

    An example of a  tensor-train (TT) layer is illustrated in Fig. \ref{fig:prior_path}.
    A linear layer with an input shape of $[128, 768]$ and a weight shape of $[768, 768]$ is represented in the TT format.
    Here the tensorized shapes are $M_i = [12, 8, 8]$, $N_i = [8, 8, 12]$, and $R_i = [1, 8, 8, 8, 8, 8, 1]$.
    Two contraction sequences are visualized as computation graphs. Notably, \textbf{Scheme-2} involves significantly more FLOPs and memory accesses compared to \textbf{Scheme-1}, highlighting the impact of execution sequence on computational efficiency.

    \begin{table*}[]
      \centering
      \caption{Feature comparison: FETTA \textit{vs.} training accelerators and TNN inference accelerators}
      \label{tab:acc_feature}
      \vspace{-5pt}
      \begin{threeparttable}
        \renewcommand{\arraystretch}{1.0}
        \resizebox{0.85\linewidth}{!}{
          \begin{tabular}{l|llllccl}
            \toprule
            \multicolumn{2}{c}{\multirow{2}{*}{Accelerators}}                    &             \multicolumn{2}{c}{Dataflow}               & \multirow{2}{*}{Data Layout Reordering}    & \multirow{2}{*}{Scenarios}      & \multirow{2}{*}{TNN Support}      \\ \cline{3-4}
            \multicolumn{2}{l}{}                                                 & Loop Ordering             &  Loop Parallelism          &                                            &                                         &                                   \\ \midrule
            \multicolumn{2}{c}{Rapid \cite{rapid}}      &  WS                       & Fixed                      & Special function units                     & {Training} & {\large \ding{55} } & \\
            \multicolumn{2}{c}{FAST \cite{zhang2022fast}}                  & WS in FP/BP, OS in WG     & Fixed                      & Transposable systolic array                & {Training} & {\large \ding{55}   } & \\
            \multicolumn{2}{c}{TRETA \cite{shao2023treta}}                       & WS, OS                    & Flexible                   & Off-Chip                                   & {Training}                          & {\large \ding{55}} &                         \\
            \multicolumn{2}{c}{SIGMA \cite{qin2020sigma}}                        & WS, IS, NLR;              & Flexible                   & Off-Chip                                   & {Training}                          & {\large \ding{55}} &                         \\ \midrule

            \multicolumn{2}{c}{ETTE  \cite{gong2023ette}}                        &  Look-ahead style         & Fixed                      & Dedicated memory access                 & {Inference}             & TT Only  &                         \\
            \multicolumn{2}{c}{Tetrix \cite{zhang2024tetrix}}                    &   WS, OS                  & Fixed                      & Dataflow switch and  Special units      & {Inference}           & All  &                         \\ \midrule
            \multicolumn{2}{c}{FETTA }                    & WS, IS, OS                & Flexible                   & \begin{tabular}[c]{@{}l@{}} Transposable systolic array, \\ Hierarchical structure,\\ Distribution and reduction network \end{tabular}  & {Training}& All  \\ \bottomrule
          \end{tabular}
        }
        \begin{tablenotes}
          \footnotesize
          \item[*] WS : weight stationary; IS: input stationary; OS: output stationary; NLR: no local reuse. \cite{chen2016eyeriss}
      \end{tablenotes}
      \end{threeparttable}
      \vspace{-5pt}
    \end{table*}

    However, existing researches usually use a fixed contraction sequence for tensorized layers. T3f \cite{t3f} and tensorly \cite{tensorly} libraries opted to reconstruct the original weight matrix from core tensors and then process it as a standard neural layer, following \textbf{Scheme-2} in Fig. \ref{fig:prior_path}.
    TIE \cite{deng2019tie} and ETTE \cite{gong2023ette} proposed computing schemes for TT and TTM format, respectively, that perform tensor contraction operations in ascending order of the core tensor index,  corresponding to \textbf{Scheme-1} in Fig. \ref{fig:prior_path}.

    Tetrix \cite{zhang2024tetrix} introduced a breadth-first approach to identify the optimal contraction sequence for TNN inference by defining a search space with $O(K!)$ candidates.  However, Tetrix treats the input node $\ten{X}$ as a fixed starting point and iteratively merges it with connected weight nodes, thereby limiting the overall size of the search space.
    Although this approach performs well for inference, there is a significant impact on training performance. Especially when the batch size dimension $B$ is large, it appears in every contraction step, leading to increased computational and memory requirements.

    \begin{figure}[tbp]
      \centering
      \includegraphics[width=0.95\columnwidth]{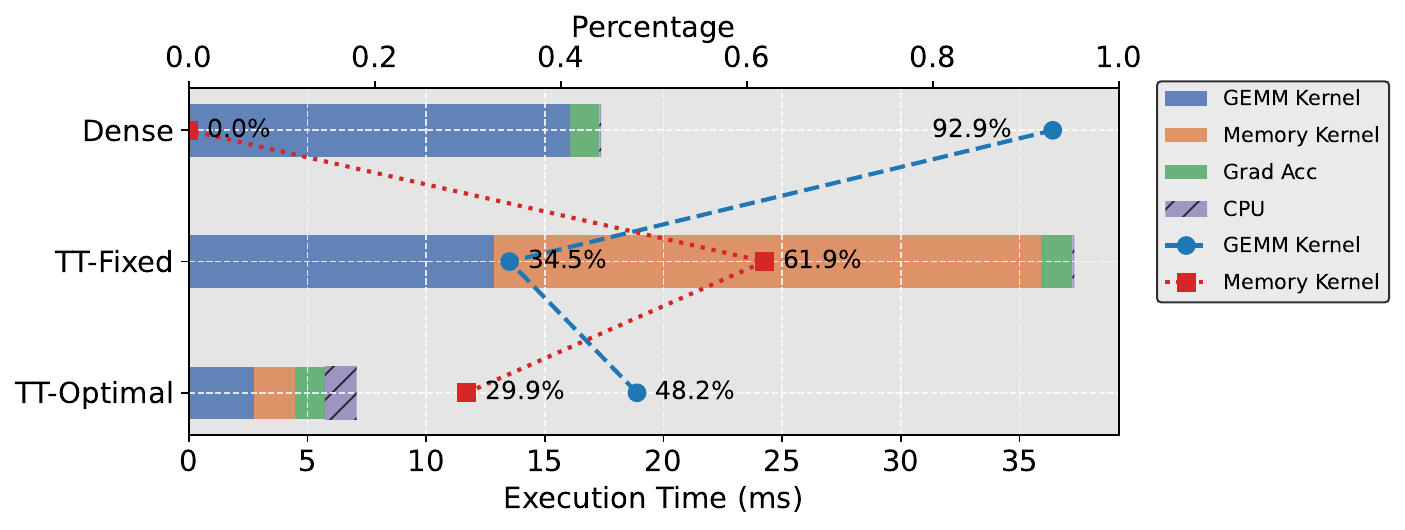}
      \vspace{-10pt}
      \caption{Training Performance Profile on GPU.}
      \label{fig:gpu_prof}
      \vspace{-15pt}

    \end{figure}

    {\bf Proposed Approach:} We propose a contraction sequence search engine (CSSE) to determine the optimal path to implement a hardware-friendly computing scheme tailored for TNN training.
    We create an enlarged search space for contraction sequences by allowing  multiple-source contractions.
    A two-stage search strategy is then employed to identify the most efficient sequence within this enlarged search space.
    The resulting contraction sequence maximizes computational efficiency and is tailored for the unique demands of TNN training.

    \subsection{Hardware Design Consideration}\label{sec:hw_consider}
    Even with an optimal computational sequence, achieving ideal performance on real hardware platforms is far from guaranteed. The practical challenges are often more complex.
    There are many accelerators for standard training \cite{zhang2022fast, rapid, zhao2021cambricon, qin2020sigma, shao2023treta} and TNN inference \cite{gong2023ette, zhang2024tetrix,deng2019tie,gong2022algorithm} , but none are specifically designed for TNN training.

    \subsubsection{Tensorized Layer on GPU}
    {
      We profiled the GPU activity during the training of with dense and TT layers, respectively. As illustrated in Fig. \ref{fig:gpu_prof}, the GPU efficiently manages data layout reordering during the training of the dense layer. Consequently, the General Matrix Multiplication (GEMM) kernel accounts for 92.9\% of the total execution time. In this process, CUDA implicitly performs layout reordering by adjusting the stride, eliminating the need for additional memory operations.

      However, in the case of the tensorized layer executed under \textbf{Scheme-1} (TT-Fixed), memory operations constitute 61.9\% of the execution time, resulting in an overall training duration that exceeds that of the dense layer. This inefficiency arises because CUDA is unable to handle the more complex layout reordering solely by adjusting the stride. Instead, it must explicitly copy data and generate new tensors to accommodate computational requirements.

      Even with an optimized contraction sequence (TT-Optimal), while the overall execution time is reduced, memory operations still account for 30\% of the total execution time, thereby diminishing the computational efficiency of the GPU. Consequently, the utilization of the GEMM kernel decreases to 48.2\%, preventing the tensorized layer from fully leveraging its acceleration potential.
    }
    \begin{figure}[tbp]
      \centering
      \includegraphics[width=0.95\columnwidth]{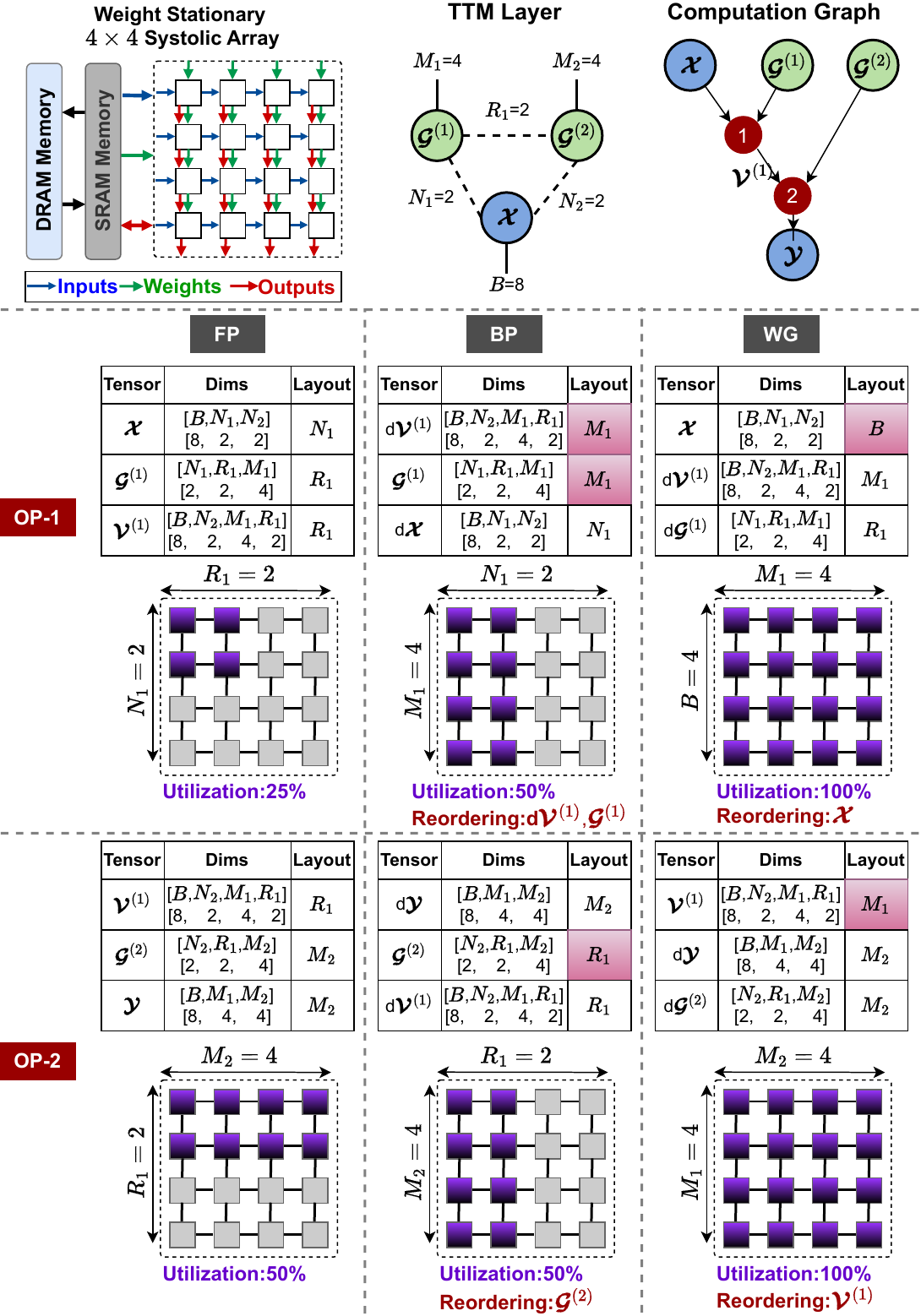}
      \caption{Training mapping of a TTM layer on a systolic array.}
      \label{fig:tpu_impl}
      \vspace{-5pt}
    \end{figure}

    \subsubsection{Tensorized Layer on TPU}
    A systolic array design is utilized in Google's tensor processing units (TPUs) \cite{jouppi2017datacenter, norrie2021design, jouppi2023tpu}
    because of its ability to efficiently handle the highly parallelized computations in matrix multiplications.
    Data stored in the off-chip memory are loaded into the on-chip SRAM memory. Input data (\textit{e.g.}, weights and activations) are fetched from SRAM and then streamed through the array in a pipelined manner, reducing memory bandwidth requirements by enabling efficient local data reuse.
    However, TPUs face under-utilization during TNN training due to their limited dataflow flexibility and the inconsistency of data layouts.

    Fig.~\ref{fig:tpu_impl} shows the execution of a 2nd-order TTM layer on a TPU-like architecture with a weight stationary systolic array of size $4\times 4$.
    The TTM layer involves two consecutive tensor contraction operations.
    Due to the lack of dataflow flexibility, mapping tensor contraction operations with small dimension sizes on a TPU causes under-utilization of computational resources. The utilization falls below 50\% during the forward and backward phases.
    The data layout is denoted as the last dimensions stored in a memory line, which indicates how many data elements can be accessed concurrently.
    As shown in Fig. \ref{fig:tpu_impl}, to achieve better utilization, a total of 5 data layout reordering operations are required to make it consistent with the dataflow. These additional reordering operations further degrade compute efficiency.
    As the number of dimensions and nodes in a tensorized layer increases, the problem becomes even more pronounced, exacerbating the inefficiencies.

    \subsubsection{Inefficiency of Previous Accelerators}
    Table~\ref{tab:acc_feature} summarizes the features of several recent accelerators designed for DNN training and TNN inference.
    RapiD \cite{rapid} adopts weight-stationary dataflow during all training phases. A specialized function unit is integrated to handle data shaping operations, which costs extra processing latency and resources.
    FAST \cite{zhang2022fast} utilizes weight-stationary and output-stationary dataflow for FP/BP and WG phases, respectively. FAST eliminates explicit matrix transpose by developing a transposable systolic array,  offering a low-complexity and low-latency solution for the different training phases of linear layers.
    TRETA \cite{shao2023treta} and SIGMA \cite{qin2020sigma} overcome the shortcomings of TPUs by enhancing the dataflow flexibility through the integration of on-chip distribution and reduction networks.
    However, they do not provide on-chip data layout transposition capabilities.
    Consequently, off-chip data layout reordering is required, leading to increased latency and energy consumption due to additional DRAM access. Alternatively, direct access to on-chip SRAM incurs the risk of bank conflicts, which can further degrade performance.

    The dataflow and architecture of ETTE \cite{gong2023ette} are dedicated to TT inference under a fixed contraction sequence, making it unsuitable for extension to training scenarios.
    Tetrix \cite{zhang2024tetrix} enables hybrid inner-outer mapping by supporting weight-stationary and output-stationary dataflow switching, eliminating the need for the most expensive transpose operation.
    However, Tetrix is designed for inference, where each intermediate tensor only needs to be used for once computation. It only supports a one-time tensor data layout transformation during the output collection/reduction stage.
    In training, tensors may be reused in different phases, each requiring distinct data layouts and dataflows.
    For example, in Fig. \ref{fig:tpu_impl}, $\ten{V}^{(1)}$ is used for FP and WG phases of \textbf{OP-2}.
    In this case, the $R_1$-last data layout is suitable for FP phase, whereas $M_1$-last data layout is suitable for WG phase, respectively. Therefore, data layout manipulations are needed not only at the output stage but also at the input stage to accommodate these differing requirements.

    \subsubsection{{\bf Proposed Implementation}}
    To overcome the inefficiency of prior works, this work proposes a~ \textbf{\underline{F}}lexible and \textbf{\underline{E}}fficient \textbf{\underline{T}}ensorized \textbf{\underline{T}}raining \textbf{\underline{A}}ccelerator (\textit{FETTA}), designed based on a hierarchical transposable systolic array.
    Simultaneously, symmetrical transposable butterfly networks are incorporated for data distribution and reduction.
    Our design provides the following key features:
    \begin{itemize}[leftmargin=*]
      \item[\ding{202}] flexible dataflow in loop ordering (IS, WS, and OS) that maximizes data local reuse across different training phases.
      \item[\ding{203}] flexible dataflow in loop parallelism that facilities tensor contraction operations across a wide range of tensor orders and dimension sizes.
      \item[\ding{204}] implicit data layout reordering during computation to optimally meet dataflow requirements.
    \end{itemize}
    These features work together to optimize computational efficiency and boost system performance.

    \section{TNN Computing Scheme}\label{sec:algorithm}
    \subsection{Contraction Sequence Search Engine}

    \textbf{Search Space:} To provide an exhaustive search for tensor contraction sequences, we allow all possible contraction orders within a tensorized layer.
    For a tensor network diagram with $K$ nodes, tensor contraction can occur between any pair of nodes.
    Each time a contraction is performed, the two nodes are merged into a new node, reducing the tensor graph to ($K-1$) nodes.
    Consequently, the entire search space has $\mathcal{O}(\prod _{i=2}^{K}C(i, 2))$\footnote{$C(n, k)$ is the combination number, which equals to $ \frac{n!}{k!(n-k)!}$.} possible contraction sequences.
    {
    Even though we enlarge the search space against Tetrix's $\mathcal{O}(K!)$, as the value of $K$ is usually less than 10,
    the additional searching cost is still reasonable. In addition, the searching is only performed once for each model, and the overhead can be amortized over the training process, thus it can be ignored compared with the overall training time.
}

    \begin{algorithm}[t]
      \caption{Contraction Sequence Search Engine}\label{alg:path_find}
      \begin{algorithmic}[1]
      \State \textbf{Input:} Input tensors in a tensorized layer: $\ten{X}, \ten{G}^{(1)},... \ten{G}^{(d)}$
      \State \textbf{Output:} Optimal contraction sequence $Best\_Seq$.
      \State $G(V, E) \leftarrow \{\ten{X}, \ten{G}^{(1)},... \ten{G}^{(d)}\}$  \Comment{{\color[HTML]{38b000}\textbf{Initialize}}}
      \State $Best\_Cost \leftarrow \infty$
      \State $Best\_Seq \leftarrow [\, ] $
      \State $Candidates \leftarrow $ a list of size N
      \State Recursive\_Search($G, 0, [\, ]$)        \Comment{{\color[HTML]{38b000}\textbf{Stage-1}}}
      \ForAll{$Seq \in Candidates$}                  \Comment{{\color[HTML]{38b000}\textbf{Stage-2}}}
          \State $Cost \leftarrow \text{Performance\_Model}(Seq)$
          \If {$Cost < Best\_Cost$}
              \State $Best\_Cost \leftarrow Cost$
              \State $Best\_Seq \leftarrow Seq $
          \EndIf
      \EndFor
      \Statex
      \Procedure{\text{Recursive\_Search}}{$G, Acc\_FLOPs, Seq$}
      \If{len($V$) == 1  and $Acc\_FLOPs < Candidates.max() $}
          \State $Candidates.\text{insert}(\{Seq, Acc\_FLOPs \})$
      \EndIf
      \ForAll{$ \{v_i, v_j \} \in V $}
        \State $v_k \leftarrow v_i  v_j$
        \State $Acc\_FLOPs \leftarrow Acc\_FLOPs + \text{FLOPs}(v_i, v_j, v_k)$
        \State $V' \leftarrow V \setminus  \{v_i, v_j\}$
        \State $V' \leftarrow V' \cup  \{v_k \}$
        \State Create graph: $G'(V', E')$
        \State $Seq \leftarrow Seq.\text{append}(\{v_i, v_j \})$
        \State Recursive\_Search($G', Acc\_FLOPs, Seq$)
      \EndFor
      \EndProcedure
    \end{algorithmic}
  \end{algorithm}

    \textbf{Cost Predictor:} Prior works primarily use the number of FLOPs as a metric of computation cost, which is intuitive and easy to calculate.
    To more precisely reflect runtime hardware performance, an analytical performance model is introduced \cite{zigzag}.
    The performance model evaluates accurate hardware performance metrics, such as latency and energy. Additionally, it performs exhaustive design space exploration to identify the optimal dataflow mapping strategy for each tensorized layer.

    \textbf{Search Engine:}  However, searching for both contraction sequences and dataflow can be highly time-consuming.  To address this, we propose a two-stage search engine to reduce the search budget. The detailed process of the contraction sequence search engine (CSSE) is elaborated in Algorithm \ref{alg:path_find}.
    \begin{itemize}[leftmargin=*]
      \item[\ding{202}] \textbf{Initialize}:  Given a tensorized layer, the corresponding tensor network diagram $G(V, E)$ is first constructed, where $V$ is the set of tensor nodes and $E$ is the set of connected edges. The best sequence ($Best\_Seq$) and the best cost value ($Best\_Cost$) are initialized as an empty list and an infinite value, respectively. A list of $Candidates$  with a size $N$ is also initialized.
      \item[\ding{203}] \textbf{Stage-1}: A depth-first search procedure is called iteratively on the tensor network diagram, with the number of FLOPs used as the cost predictor.
      In each iteration, the current diagram $G(V, E)$, the cumulative contraction cost ($Total\_FLOPs$), and the recorded sequence ($Seq$) are used as inputs.
      \begin{itemize}[leftmargin=*]
        \item[a)] Enumerate all possible pairs of nodes ${v_i, v_j}$ in the graph.
        \item[b)] Perform a contraction operation between $v_i$ and $v_j$, resulting in a new node $v_k$. Calculate the contraction cost and add it to $Total\_FLOPs$.
        \item[c)] Update the graph to $G'(V', E')$ by removing the nodes ${v_i, v_j}$ and adding the newly generated node $v_k$.
        \item[d)] Append the contracted pair ${v_i, v_j}$ to the sequence $Seq$.
        \item[e)] Pass the updated graph, total cost, and sequence to the next level of the search.
      \end{itemize}
      When the graph contains only one node, the search function has reached the end of a contraction sequence. At this point, the candidate list is updated by comparing the current sequence and cost with existing candidates.
      \item[\ding{204}] \textbf{Stage-2}: all candidate sequences are evaluated by the performance model.
      The performance model determines the best sequence based on the desired hardware performance metric, such as latency, energy consumption, or energy-delay product (EDP). The optimal sequence ($Best\_Seq$) is then selected to ensure maximum computational efficiency.
    \end{itemize}

    \section{FETTA Architecture}\label{sec:architecture}

    \subsection{Overview}\label{sec:hw_overview}
\begin{figure}[t]
      \centering
      \includegraphics[width=0.95\linewidth]{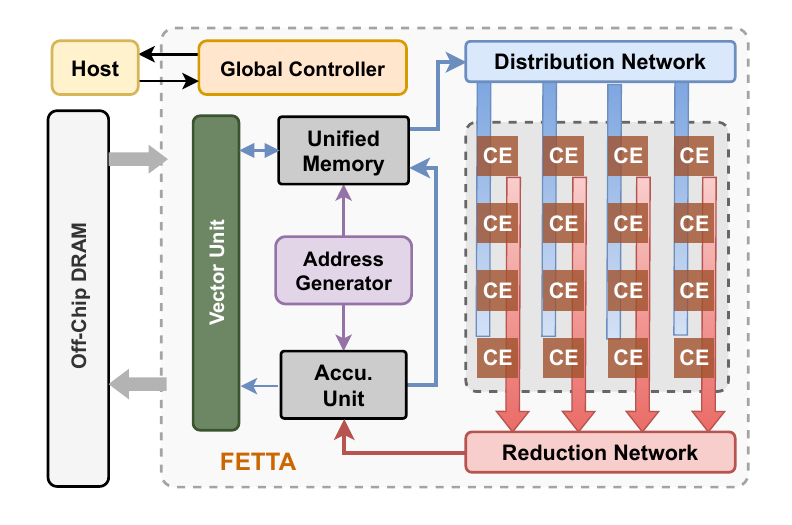}
      \vspace{-5pt}
      \caption{Overview of FETTA system architecture.}
      \label{fig:overall}

    \end{figure}

    Fig. \ref{fig:overall} illustrates the overall architecture of FETTA, which comprises a global controller, a multi-banked unified memory, a tensor contraction unit (TCU), a vector unit, an accumulation unit, and an address generator for on-chip memory access.

    The controller decodes incoming instructions, configures dataflow, and coordinates the operations of the subsequent components during different training phases.
    A unified memory is utilized to store activations, weights, and gradients, offering flexibility for complex operations in training.
    Unlike inference with two operands of input activations and weights, the training process also involves gradients. Therefore, the on-chip memory must be dynamically allocated for tensors during different training phases.

    TCU is developed to perform tensor contraction operations in tensorized training. It comprises a processing element (PE) array organized in a hierarchical structure.
    The address generator generates memory addresses, and the fetched data are dispatched to the PE array via a distribution network.
    The PE array performs tensor computations with a flexible dataflow and sends the results to the accumulator through a reduction network. Upon completion of a tensor contraction operation, the results are written back to the unified memory or external DRAM. If required, a vector unit is used to process non-linear functions.
    \begin{figure}[t]
      \centering
      \includegraphics[width=0.85\linewidth]{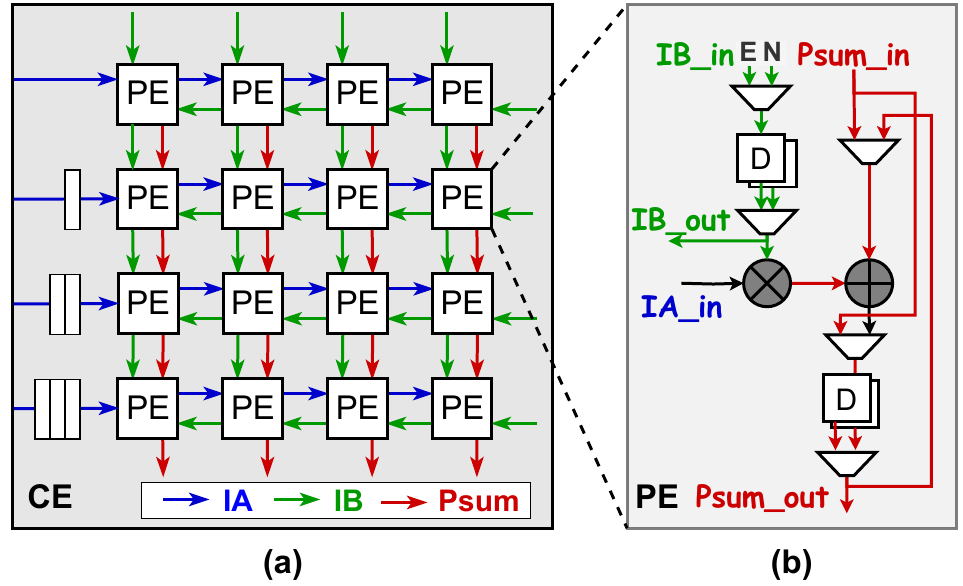}
      \vspace{-8pt}
      \caption{Micro-architectures of (a) Contraction Engine and (b) Processing Element.}
      \vspace{-8pt}
      \label{fig:ce_pe}
    \end{figure}

    \begin{figure*}[t]
      \centering
      \includegraphics[width=0.95\linewidth]{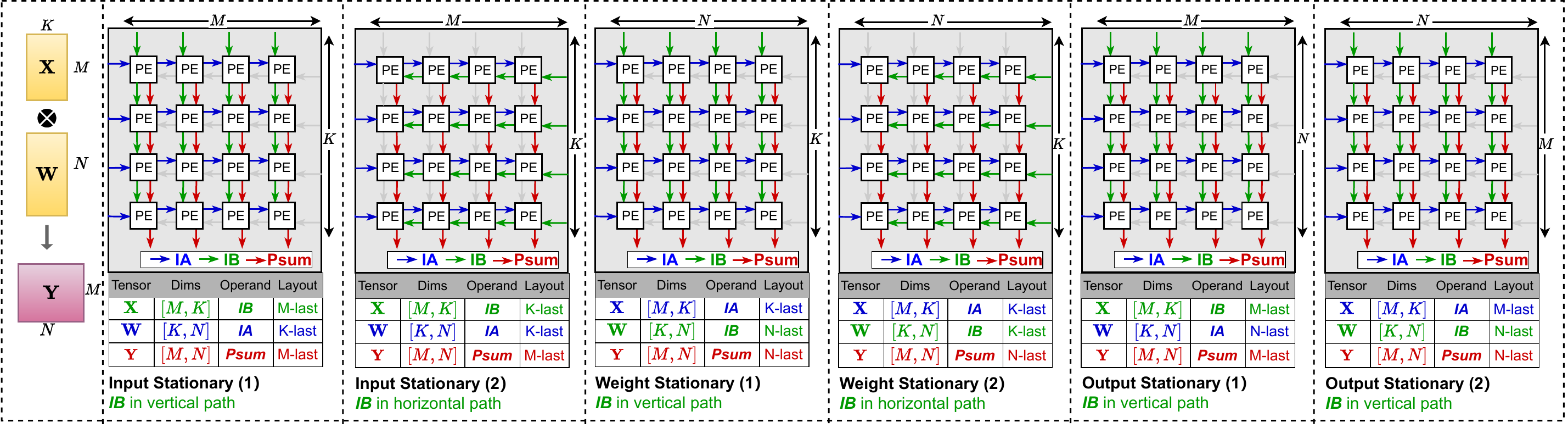}
      \vspace{-8pt}
      \caption{Illustration of dataflow modes supported by the Contraction Engine.}

      \label{fig:ce_mode}
    \end{figure*}
    \subsection{Tensor Contraction Unit}
    To enable a flexible dataflow that maximizes data parallelism and reuse, we design a tensor contraction unit (TCU). The TCU consists of 16 contraction engines (CEs), which are connected to the unified memory through a distribution network, and to the accumulation unit through a reduction network.

    \subsubsection{Contraction Engine}
    The micro-architecture of a CE is presented in Fig. \ref{fig:ce_pe}{(a)}. To facilitate diverse dataflows and computational patterns during different training phases, the CE is designed as a reconfigurable and transposable systolic array with of size $4\times 4$. The micro-architecture of each PE is shown in Fig. \ref{fig:ce_pe}{(b)}.
    Since different data types are involved in different training phases, we avoid using the term \textit{input activations} and \textit{weights} to represent operands at the architectural level.
    Instead, in a tensor contraction operation, two input operands are referred to as \textit{IA} and \textit{IB}, while the output operand is denoted as partial sum (\textit{Psum}).
    Depending on the training phase or dataflow, either \textit{IA} or \textit{IB} can be associated with input activations ($\mat{X}$), weights ($\mat{W}$), or output gradients (d$\mat{Y}$).
    Similarly, \textit{Psum} can serve for output activations ($\mat{Y}$), input gradients (d$\mat{X}$), or weight gradients (d$\mat{W}$).

    As shown in Fig. \ref{fig:ce_pe}{(a)}, \textit{IAs} are horizontally sent to the PEs in the same row from the left.
    Given the relatively small size of the CE, \textit{IAs} can be broadcast simultaneously to all PEs in the same row without streaming between registers. Besides, \textit{IAs} in different rows are skewed to ensure functional correctness.
    The datapath for \textit{IB} is transposable and reconfigurable, which allows \textit{IBs} to enter the PE array either horizontally or vertically depending on the required dataflow.
    Furthermore, based on the architecture of PE in Fig. \ref{fig:ce_pe}{(b)}, the operand represented by \textit{IB} can be held stationary to support  input-stationary and weight-stationary dataflows.
    In these modes, \textit{IBs} are held in PE registers and  reused for multiplications with different \textit{IAs}, while \textit{Psums} are accumulated along the column direction and streamed out from the bottom side of the array.
    When the CE operates in output-stationary dataflow mode, \textit{IBs} are streamed vertically into the PE array from the top, and \textit{Psums} are locally accumulated within each PE. Once the final results are obtained, \textit{Psums} are streamed out from the bottom and written back to memory.
    To hide bubbles due to the pre-loading of \textit{IBs}  and the streaming of \textit{Psums}, a double buffering technique is implemented for \textit{IB} and \textit{Psum} registers.

    Fig. \ref{fig:ce_mode} illustrates an example of executing a GEMM operation on a contraction engine. Depending on the selected dataflow, the datapath of \textit{IB}, and the tensor-operand relations, there are six feasible mapping strategies, which also indicate different data layouts for tensors.
    For higher-order tensor contraction operations, the number of feasible dataflow choices increases, highlighting the flexibility of the intra-CE level. The optimal dataflow can be identified through exhaustive evaluation and search.

    \subsubsection{Distribution Network}
\begin{figure*}[tbp]
      \vspace{-10pt}
      \centering
      \subfigure[]{
        \includegraphics[height=0.24\textwidth]{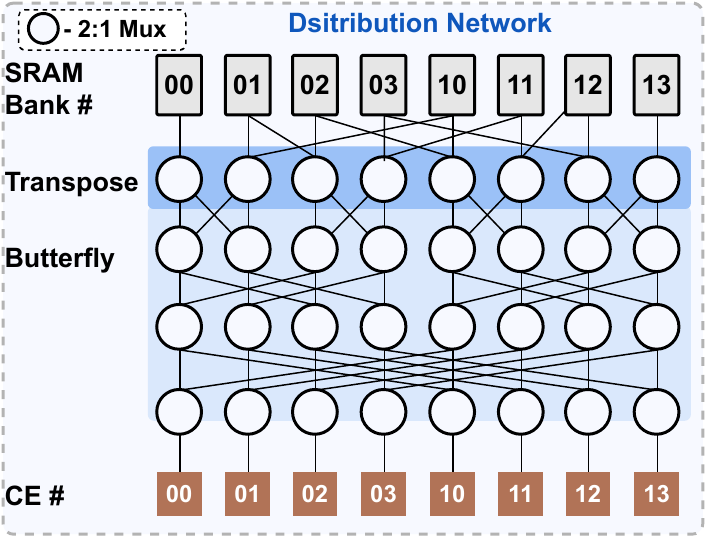}
        \label{subfig:distnet}
      }
      \hspace{10pt}
      \subfigure[]{
        \includegraphics[height=0.24\textwidth]{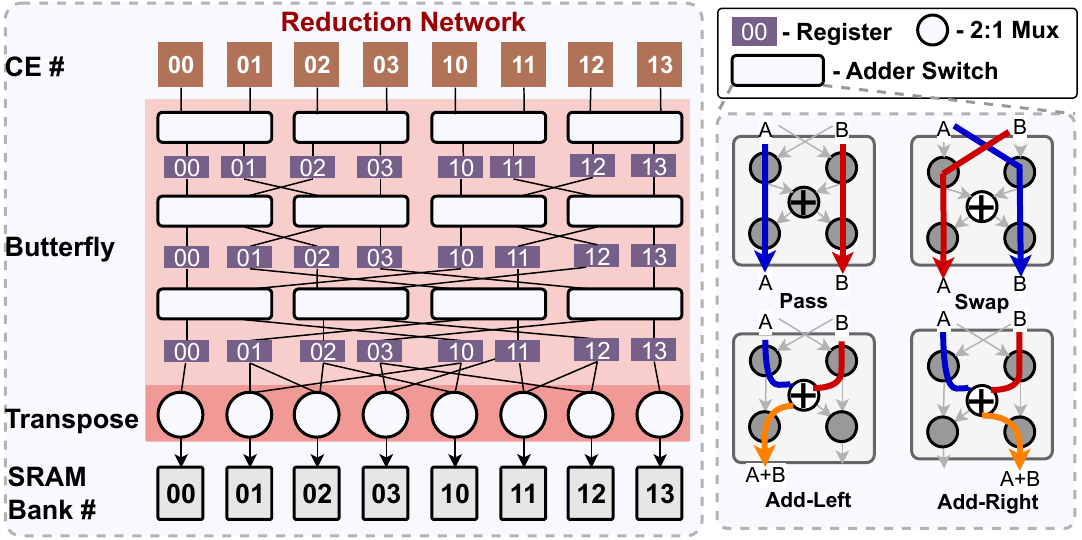}
        \label{subfig:redunet}
      }
      \vspace{-5pt}
      \caption{Micro-architectures of (a) Distribution Network and (b) Reduction Network.}

      \vspace{-10pt}
      \label{fig:noc}
    \end{figure*}

    A distribution network delivers the data read from the unified memory to the CE array, as illustrated in Fig. \ref{subfig:distnet}.
    A transposable butterfly network is employed to provide dataflow flexibility at the inter-CE level.
    The transposable butterfly is an N-input N-output multi-stage network with $log(N)+1$ levels, which enhances the traditional butterfly topology \cite{tong2024feather} by stacking a transpose layer at the first level. Each level consists of $N$ 2:1 multiplexers (Mux), each controlled by a $1-bit$ signal to decide whether the output is derived  from the vertical or diagonal input.

\begin{figure}[tbp]
      \centering
      \includegraphics[width=0.9\linewidth]{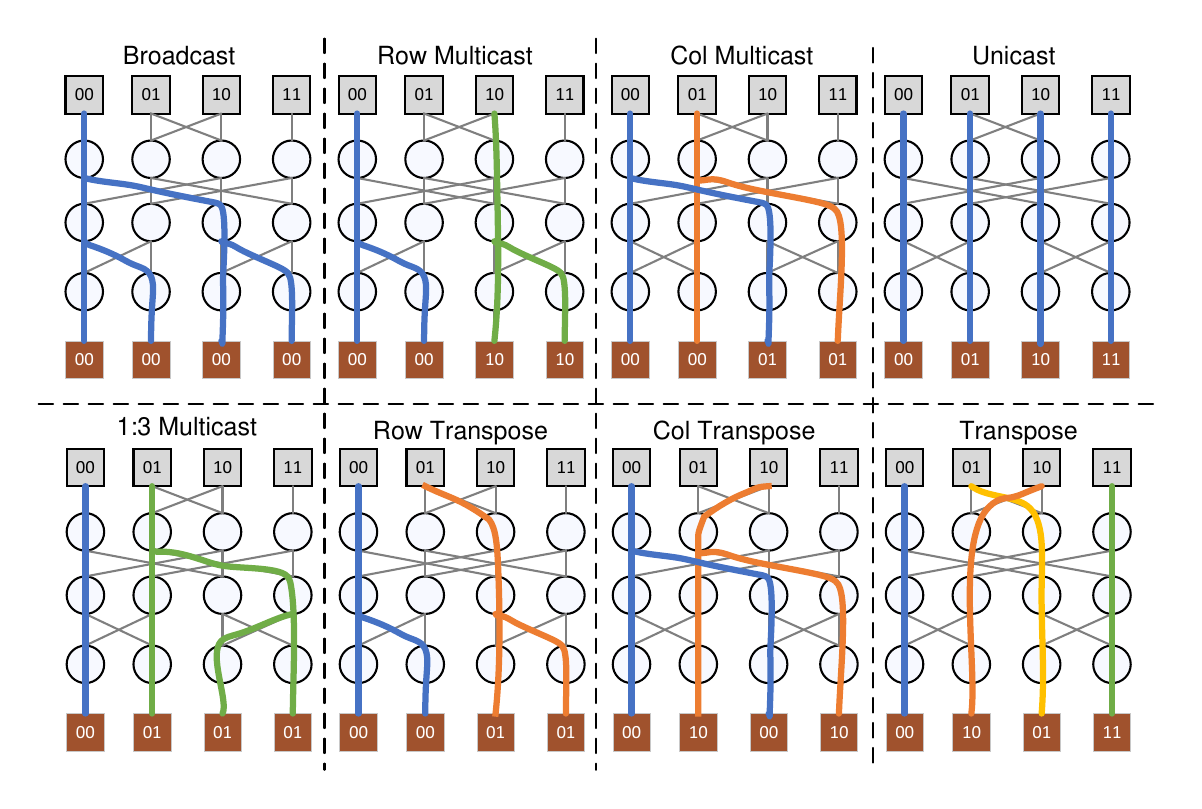}
      \vspace{-8pt}
      \caption{Examples of working modes in distribution network.}
      \vspace{-10pt}
      \label{fig:dist_mode}
    \end{figure}

    The transposable butterfly network is a blocking network that is not designed for arbitrary reordering without congestion, as is possible with crossbar and Benes networks \cite{arora1990line, qin2020sigma}.
    However, it provides sufficient flexibility for unicast (one-to-one) and various multicast (one-to-many) with transposable capability, as shown in Fig. \ref{fig:dist_mode}.
    Compared with crossbar and benes networks, which have hardware complexities of $N^2$ and $2N log(N)$, respectively, our design achieves a better balance between hardware complexity and flexibility, making it well-suited for tensorized neural network training.

    \subsubsection{Reduction Network}
    A reduction network receives the output \textit{Psums} from the CE array and transmits them to the accumulation unit.
    As shown in Fig. \ref{subfig:redunet}, the reduction network is also built with a transposable butterfly topology.
    Unlike the distribution network, the reduction network not only facilitates data transfer between the CE array and the accumulator array, but also performs spatial reduction across different CEs.
    Specifically, the reduction network is an N-input N-output multi-stage network with $log(N)+1$ levels, enhanced with a transpose layer attached at the bottom of the butterfly network. At each level of the butterfly, there are $(N/2)$ 2-input 2-output adder switches \cite{tong2024feather}.

    The micro-architecture of an adder switch is depicted in Fig. \ref{fig:noc}, which is controlled by a 2-bit signal, enabling four distinct operational modes:
    \begin{itemize}[leftmargin=*]
      \item \textbf{Pass  / Swap }: The switch either directly passes  inputs to the output ports or swaps inputs between the output ports.
      \item \textbf{Add-Left/ Add-Right}: The switch sums the data from the input ports and transmits the result to either the left or right output port.
    \end{itemize}
    Registers placed between adjacent butterfly levels help to reduce the critical path length, ensuring timing closure.

    \subsubsection{Microarchitectural Benefits}
    By combining intra-CE flexibility through transposable PE arrays and inter-CE flexibility through distribution and reduction networks, the TCU is capable of performing training of tensorized neural networks with high utilization and efficiency.
    The TCU integrates complex data reordering operations into the computation, avoiding the implementation of dedicated memory processing units.

    SIGMA \cite{qin2020sigma} and FEATHER \cite{tong2024feather} adopted the Benes network \cite{arora1990line} for the distribution network and the reduction network, respectively.
    The Benes network provides arbitrary non-blocking data reordering by stacking two butterfly networks back-to-back, but this design doubles the area cost compared to a single butterfly network.
    As discussed in Section~\ref{sec:hw_consider}, reordering the data layout for only one operand is insufficient to meet the demands of training scenarios. For instance, SIGMA performs reordering for inputs and weights, while FEATHER focuses on outputs.
    In contrast, our design supports data reordering for both input and output operands while employing a simpler topology, achieving greater efficiency and flexibility.

    The systolic array has a compact architecture but limited flexibility, whereas the butterfly network provides greater flexibility at the expense of a hardware cost of $O(N log(N))$.
    FETTA adopts a hybrid architecture: a systolic array is used within each CE to ensure compactness, while a butterfly network is utilized across CEs to provide flexibility. Consequently, the proposed design achieves an effective trade-off between performance and hardware cost.

    \subsection{Memory Management}

    \subsubsection{Unified Memory}
    To accommodate diverse computational characteristics and corresponding data allocation patterns, as discussed in Section \ref{sec:hw_overview}, a unified memory is designed to store all types of on-chip data, including activations, weights, and gradients.
    The unified memory is physically implemented as 16 separate SRAM banks, providing simultaneous data ports for both IA and IB. A ping-pong buffer is integrated to achieve (1) latency hiding during the fetching of the next tile from off-chip DRAM, and (2) on-chip inter-layer pipelining.

    Each memory bank stores four data elements per row, matching the size of a CE. Consequently, up to $4\times 16 =64$ data elements can be fetched to the TCU per cycle, depending on the dataflow requirements.
    Since the amount of data required varies with different dataflows, the flexible memory and on-chip network design allow for adaptive adjustment of activated memory banks, avoiding unnecessary and redundant memory access.

    \subsubsection{Accumulation Unit}
    The accumulation unit temporarily stores \textit{Psums} streamed from the reduction network, particularly when the reduction size of workloads exceeds the overall reduction capacity of the TCU, as often occurs with IS and WS dataflows.
    The accumulation unit comprises 16 SRAM banks, each associated with one output port of the reduction network.
    Each bank stores four \textit{Psum} elements per row and is equipped with four corresponding adders.

    \subsection{Control Mechanism}

    \begin{figure}[t]
      \centering
      \includegraphics[width=\linewidth]{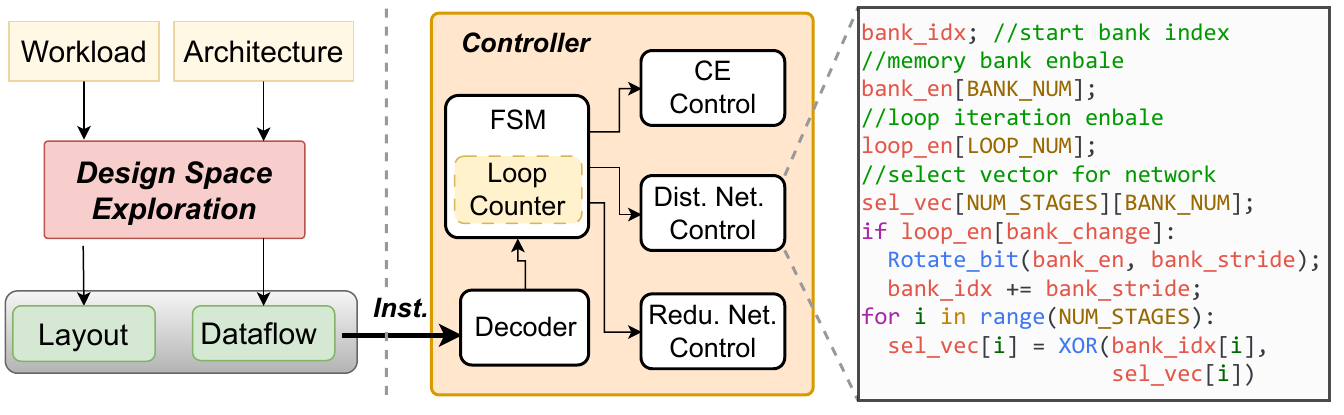}
      \vspace{-10pt}
      \caption{Execution and controller mechanism}
      \vspace{-10pt}
      \label{fig:control}
    \end{figure}
    {
    The execution and control mechanism of FEETA is illustrated in the Fig. \ref{fig:control}. Following the Design Space Exploration (DSE) phase, the dataflow and layout configurations for each workload are determined. These configurations are then translated into input instructions, which the host writes to the controller to initiate execution.

    Within the controller, the decoder is responsible for parsing the instructions, extracting the dataflow mode, and identifying the distribution/reduction working modes. This process also determines the default selection vector of networks (\textit{sel\_vec}). A finite state machine (FSM) monitors and regulates the execution flow through loop counters.

    The CE datapath is determined based on the selected dataflow mode.
    Memory bank transitions occur when execution reaches specific loop dimensions, requiring updates to the bank enable and bank index signals. Due to changes in data sources, the distribution/reduction network must be dynamically reconfigured to establish new routing paths. The control mechanism for the distribution network is depicted on the right side of Fig. \ref{fig:control}. At each stage, an XOR operation is performed on the corresponding bits of the bank index and \textit{sel\_vec}, generating the necessary control signals for reconfiguration. Similar mechanism is applied for reduction network.

    This control framework ensures efficient processing by dynamically adapting the dataflow, memory access patterns, and network configurations throughout execution.
    }

    \begin{table}[]
      \centering
      \caption{{Evaluation Benchmarks.}}
      \label{tab:benchmarks}
      \begin{threeparttable}
      \resizebox{0.9\columnwidth}{!}{
      \begin{tabular}{c|clc}
      \toprule
      \rowcolor{gray!50} \textbf{Task}                                     & \textbf{\makecell{Decomposition \\ Method}} & \textbf{\makecell{Accuracy}}   & \textbf{Params.\textdownarrow} \\ \midrule
      \multirow{2}{*}{\makecell{Transformer \\ on ATIS}}                                 & Dense  & 95.20 & -      \\
                                                                                        & \cellcolor{row_color_2} TT \cite{yang2023quantization}     & \cellcolor{row_color_2}96.00 &\cellcolor{row_color_2} $197.2\times$  \\ \midrule
      \multirow{2}{*}{\makecell{Transformer \\ on WMT14}}                               & Dense  & 34.64$^*$ & -      \\
                                                                                        & \cellcolor{row_color_2} TT     &\cellcolor{row_color_2}33.70$^*$ &\cellcolor{row_color_2} $4.3\times$    \\ \midrule
      \multirow{2}{*}{BERT on SQuAD}                                                    & Dense  & 90.68 & -      \\
                                                                                        & \cellcolor{row_color_2}TT \cite{yang2024comera}     &\cellcolor{row_color_2}88.76 &\cellcolor{row_color_2} $10.4\times$   \\ \midrule
      \multirow{5}{*}{\makecell{LSTM  on UCT-11 }}                                      & Dense  & 79.69 & -      \\
                                                                                        & \cellcolor{row_color_2} BT \cite{ye2018learning}     & \cellcolor{row_color_2}85.30 &\cellcolor{row_color_2} $17,414\times$  \\
                                                                                        & HT \cite{yin2020compressing}    & 87.20 & $47,375\times$ \\
                                                                                        &\cellcolor{row_color_2} TR \cite{pan2019compressing}    &\cellcolor{row_color_2}86.90 & \cellcolor{row_color_2} $34,193\times$  \\
                                                                                        & TTM \cite{yang2017tensor}   & 79.60  & $18,250\times$  \\ \bottomrule
      \end{tabular}
      }
      \begin{tablenotes}
        \footnotesize
        \item $*$: BELU
      \end{tablenotes}
      \end{threeparttable}
      \vspace{-15pt}
    \end{table}

      \begin{figure*}[t]
        \centering
        \hspace{-18pt}
        \subfigure[]{
          \includegraphics[width=0.19\linewidth]{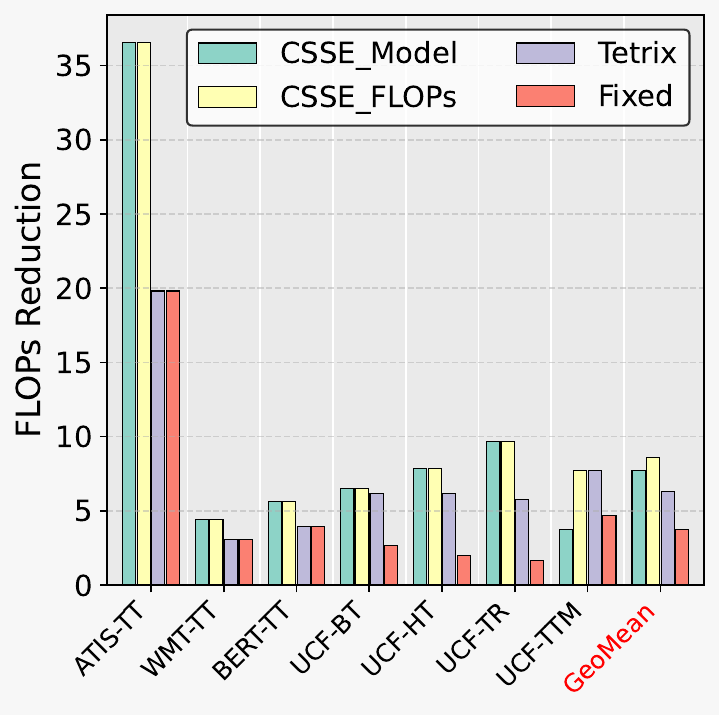}
          \label{subfig:flops}
        }
        \hspace{-10pt}
        \subfigure[]{
          \includegraphics[width=0.19\linewidth]{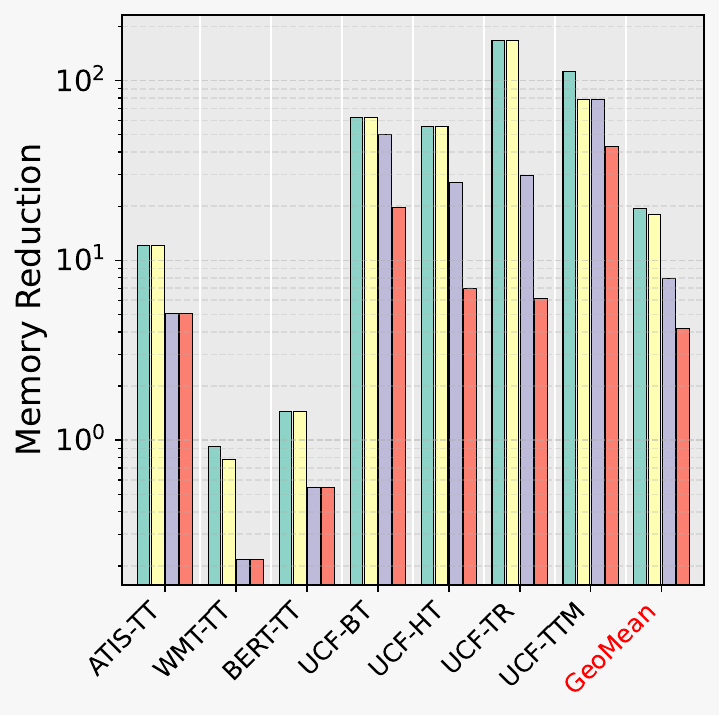}
          \label{subfig:memory}
        }
        \hspace{-10pt}
        \subfigure[]{
          \includegraphics[width=0.19\linewidth]{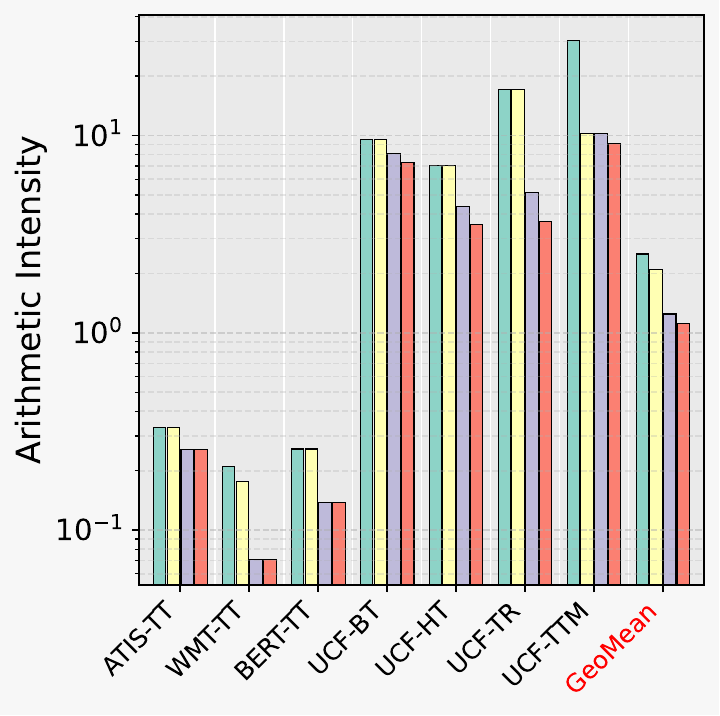}
          \label{subfig:arithmetic}
        }
        \hspace{-10pt}
        \subfigure[]{
          \includegraphics[width=0.19\linewidth]{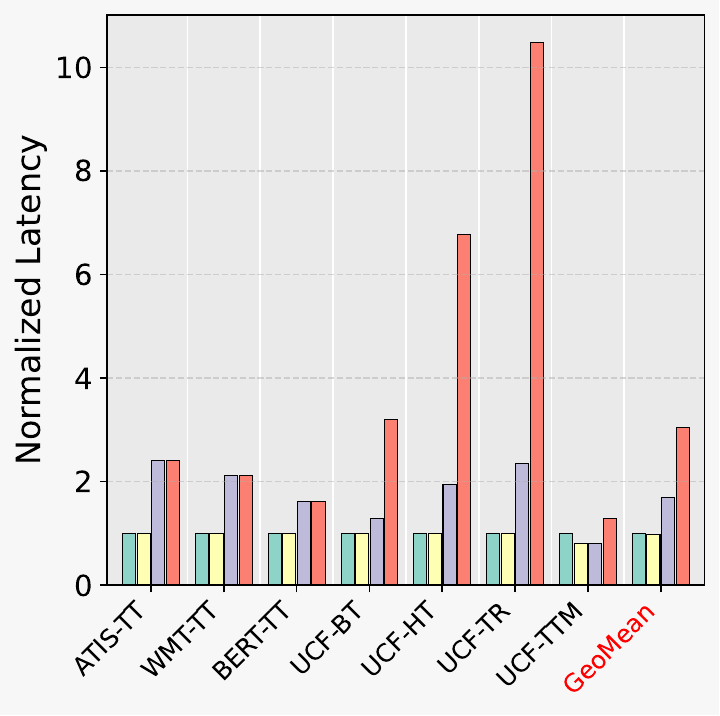}
          \label{subfig:lantency_path}
        }
        \hspace{-10pt}
        \subfigure[]{
          \includegraphics[width=0.19\linewidth]{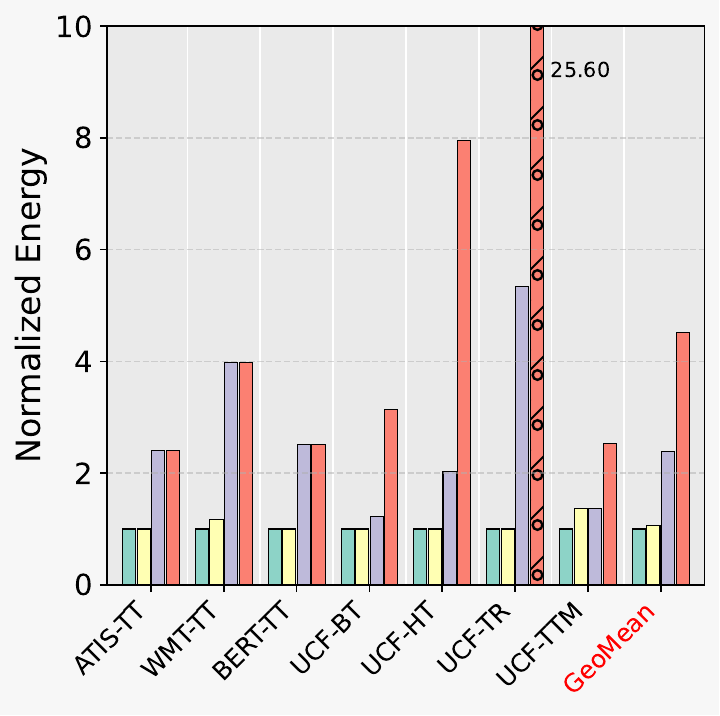}
          \label{subfig:energy_path}
        }
        \vspace{-5pt}
        \caption{Comparison of the contraction sequence search engine (CSSE) with existing strategies in terms of (a) FLOPs reduction over dense models, (b) Memory access reduction over dense models, (c) Arithmetic intensity against dense models, (d) Latency, and (e) Energy. Higher is better in (a), (b), and (c), and lower is better in (d) and (e).}

      \label{fig:path_comp}
      \end{figure*}

    \section{Evaluation Methodology}\label{sec:methods}

    \subsection{Benchmarks}
    To evaluate the performance of FETTA, we selected several models commonly used in video classification and NLP tasks,
    where different tensor decomposition methods are applied. Table~\ref{tab:benchmarks} summarizes the details of four benchmark models, including the specific decomposition methods used during training, testing accuracy, and parameter compression ratios.
    {
    The results demonstrate that tensorized training achieves accuracy comparable to, or even surpassing, the baseline while significantly reducing the number of parameters. Notably, in the UCF task, it exhibits the ability to mitigate overfitting in certain scenarios. These findings highlight the potential of TNN for enhancing the efficiency of on-device training.
    }

    \subsection{Hardware Configurations}
    FETTA consists of 16 CEs, each comprising a $4\times 4$ PE array. BFLOAT16-based multiply-accumulate (MAC) units are adopted in PEs \cite{BFLOAT16}.
    The vector unit is equipped with 64 floating-point units.
    The on-chip memory includes 512-KB SRAM in the unified memory and 128-KB SRAM in the accumulation unit. An LPDDR4 with a bandwidth of 25.6 GB/s is utilized as the off-chip memory.

    We compare FETTA against several state-of-the-art accelerators in training scenarios, including TPU v2\cite{norrie2021design}, TRETA \cite{shao2023treta}, and SIGMA \cite{qin2020sigma}.
    To ensure a fair comparison, the specifications of all baseline accelerators are aligned with  those of FETTA.  Specifically, the number of MAC units is scaled to 256, and all designs use the BF16 format. Additionally, all designs are configured to have the same total on-chip memory size.

    Besides, we compare FETTA against a general GPU, using the NVIDIA RTX 3090. Various workloads are deployed on PyTorch 2.3, CUDA 12.0. Execution time is measured by inserting { \textit{cuda.synchronize}} at the start and end points of the workload, and the elapsed time is calculated. For power measurement, power consumption is periodically recorded using the \textit{nvidia-smi} tool during runtime, and the average value is computed.

    \subsection{Simulation Infrastructure}
    The FETTA architecture was described in SystemVerilog RTL and synthesized using Synopsys Design Compiler with the ASAP 7nm PDK.
    {
      The place-and-route process was conducted with Cadence Innovus.
    }
    Power evaluation is performed using activity files generated from post-route netlist simulations and analyzed by PrimeTime PX.
    The area, power, and access latency of the on-chip memory were estimated using PCACTI.
    For off-chip DRAM memory, latency and energy consumption were estimated using the model provided by Micron.

    To accurately evaluate and analyze the performance of FETTA and prior accelerators, we further developed a cycle-accurate analytical model based on ZigZag \cite{zigzag} by integrating synthesized architecture characteristics.
    The ZigZag is enhanced to support \ding{202} tensor contraction operations and \ding{203} cross-layer data layout explorations.
    For various workloads and architectures, the optimal dataflow is identified through exhaustive design space exploration, and the final performance results are reported.

    \begin{figure}[tb]
      \centering
      \hspace{-10pt}
      \includegraphics[width=0.9\columnwidth]{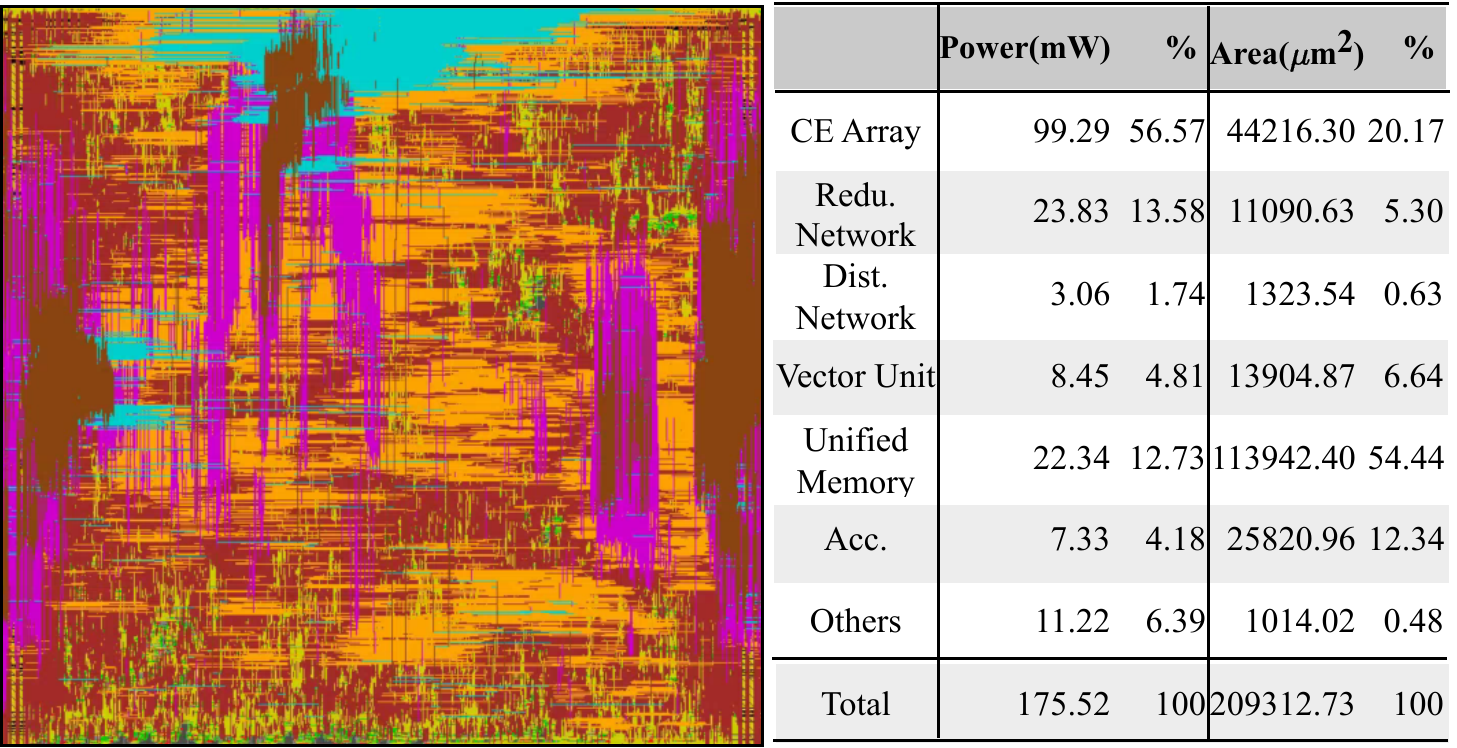}
      \caption{Layout and Hardware Characteristics of FETTA.}
      \label{fig:pnr}
      \hspace{-20pt}
    \end{figure}

    \begin{figure*}[t]
      \centering
      \hspace{-30pt}
      \subfigure[]{
        \includegraphics[width=0.5\linewidth]{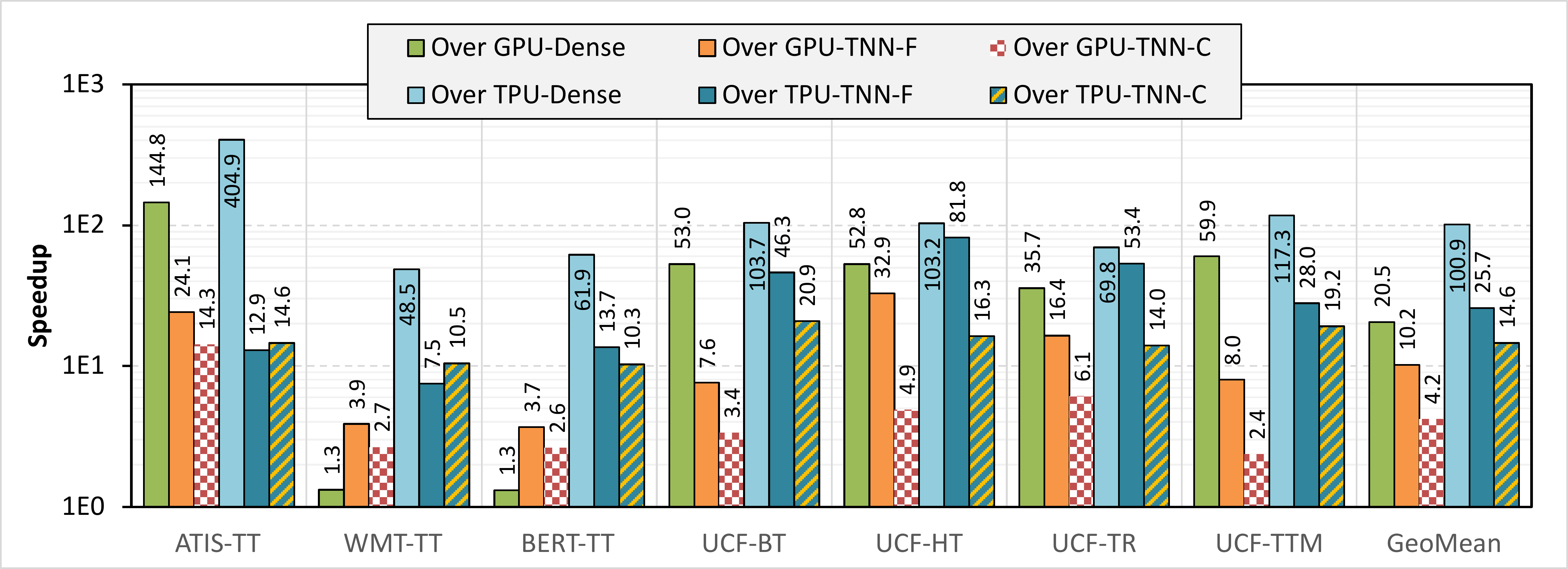}
        \label{subfig:lantency_dense}
      }
      \hspace{-5pt}
      \subfigure[]{
        \includegraphics[width=0.5\linewidth]{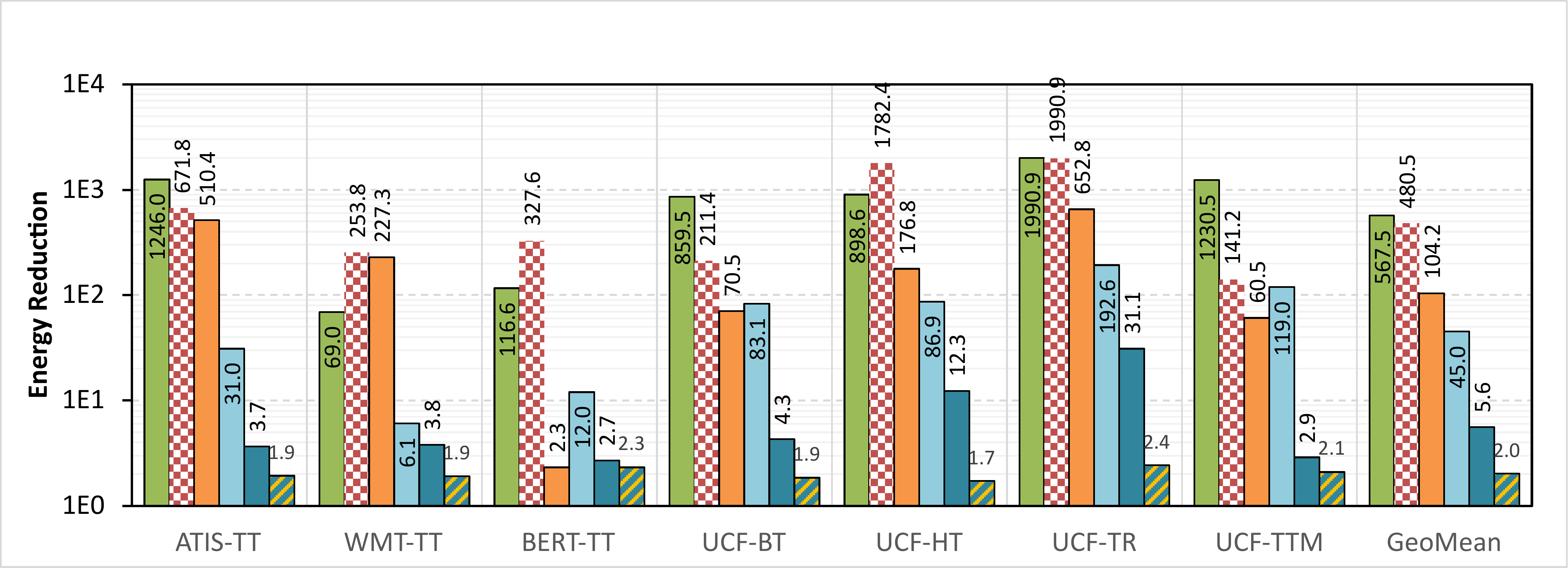}
        \label{subfig:energy_dense}
      }
      \hspace{-20pt}
       \vspace{-10pt}
      \caption{ Performance improvements of tensorized training on FETTA over GPU and TPU. (a) Speedup, (b) Energy reduction.}

      \vspace{-10pt}
      \label{fig:dense_comp}
    \end{figure*}

    \section{Evaluation Results}\label{sec:results}

    \subsection{Hardware Characteristics}

  {\color{black}
    Fig.~\ref{fig:pnr} shows the layout and the breakdown of the area and power consumption for FETTA.
    FETTA costs an area of  $209312.73 \mu m^2$ and a power of $175.52$ mW in total. FETTA operates at 1.0-GHz frequency under a supply voltage of 0.7V.
    The CE array accounts for $20.17\%$ of the area and $56.57\%$ of the power consumption.
    To enhance dataflow flexibility, the distribution and reduction networks only occupy $0.63\%/5.30\% $ of the area, and consume $1.74\%/13.58\%$ of the power respectively. The reduction network incurs higher costs than the distribution network due to the inclusion of adders and registers.}

    \subsection{Contraction Sequence Analysis}
    The improvements of the proposed contraction sequence search engine (CSSE) over the existing strategies, including the search method in Tetrix \cite{zhang2024tetrix} and fixed contraction sequences, are shown in Fig. \ref{fig:path_comp}.
    There are two variants of CSSE: CSSE-Model and CSSE-FLOPs, which take EDP from performance model and FLOPs as metrics, respectively.
    The fixed contraction sequences introduced in \cite{gong2023ette} \cite{deng2019tie}  \cite{gong2022algorithm} are applied for TTM, TT, and HT, respectively.
    Sequential contraction sequences are applied for TR and BT.
    For all contraction sequences, latency and energy results are evaluated on FETTA.

  {
    For TT-compressed models, Tetrix often follows the sequential sequence from ETTE~\cite{gong2023ette} due to a restricted search space.
    By leveraging an expanded search space, CSSE-Model identifies superior sequences, achieving a $1.42\sim 1.85 \times$ higher FLOPs reduction ratios, $1.61\sim 2.39\times$ speedup, and $2.51\sim 3.61\times$ energy reduction  compared to Tetrix.
    TT models exhibit higher arithmetic intensity than dense models due to the generation of additional intermediate tensors and the relatively lower memory reductions compared to FLOPs reductions.

    For UCF-TTM, CSSE-FLOPs and Tetrix find the same optimal sequence  due to the small node count (5 in UCF-TTM), This results in $1.65\times$ FLOPs reduction, $1.58\times$ speedup, and $1.83\times$ energy efficiency improvement compared with the fixed pattern.
    However, CSSE-Model exhibits higher FLOPs but lower memory access than CSSE-FLOPs, leading to $0.8\times$ thoughput and $1.37\times$ energy efficiency.

    For TR, which has largest number of nodes (14 in UCF-TR), CSSE significantly outperforms Tetrix and fixed sequences, demonstrating $2.07\times$/$7.38\times$ speedup and $8.49\times$/$40.64\times$ energy efficiency gains.

    On average, CSSE-Model realizes $1.22\times$/$2.07\times$ improvements in FLOPs reduction, $2.46\times$/$4.67\times$ reductions in memory access,  $1.68\times$/$3.03\times$ in speedup, and $2.38\times$/$4.52\times$ in energy efficiency compared with Tetrix and fixed sequences.

    CSSE-Model and CSSE-FLOPs occasionally yield identical paths  due to the combined influence of workload and architecture.
    Compared with CSSE-FLOPs, CSSE-Model achieves $1.10\times$ and $1.16\times$ reductions in EDP for UCF-TTM and WMT-TT, respectively.
  }

    \subsection{ Comparison with GPU and TPU}
\begin{figure}[t]
      \centering
      \hspace{-10pt}
      \subfigure[]{
        \includegraphics[width=0.24\textwidth]{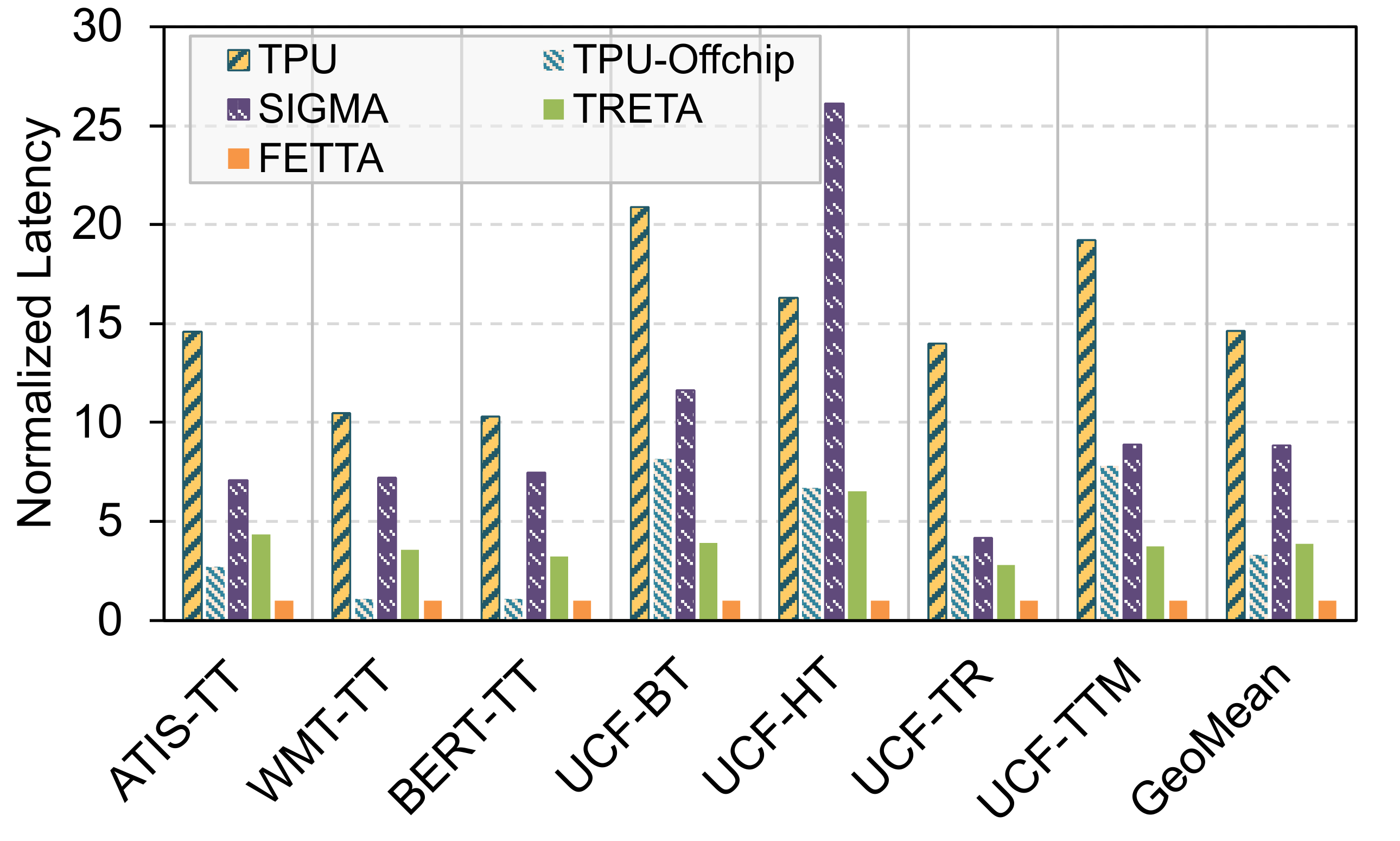}
        \label{subfig:latency_comp}
      }
      \hspace{-10pt}
      \subfigure[]{
        \includegraphics[width=0.24\textwidth]{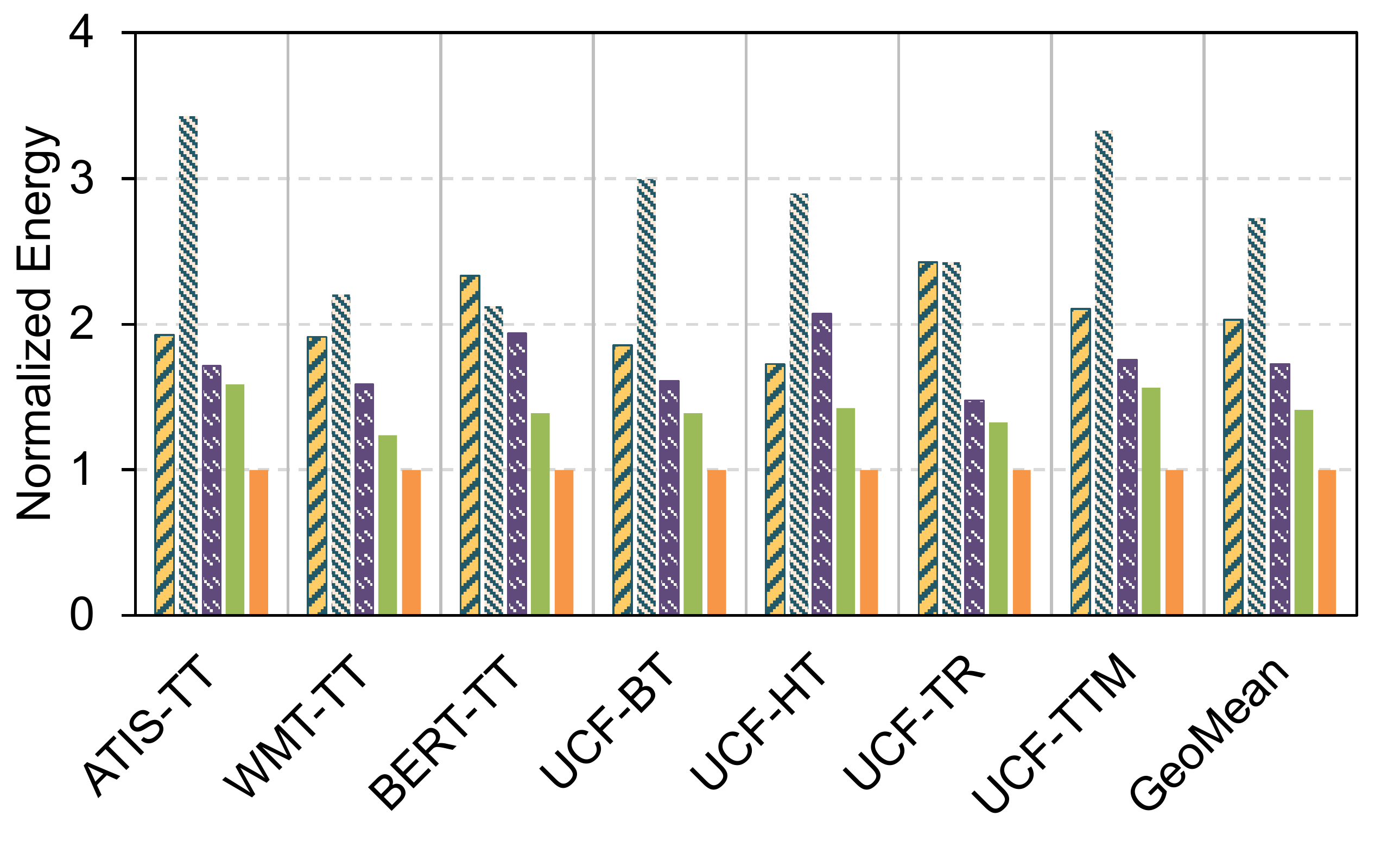}
        \label{subfig:energy_comp}
      }
      \vspace{-10pt}
      \hspace{-10pt}
      \subfigure[]{
      \vspace{-10pt}
      \includegraphics[width=0.24\textwidth]{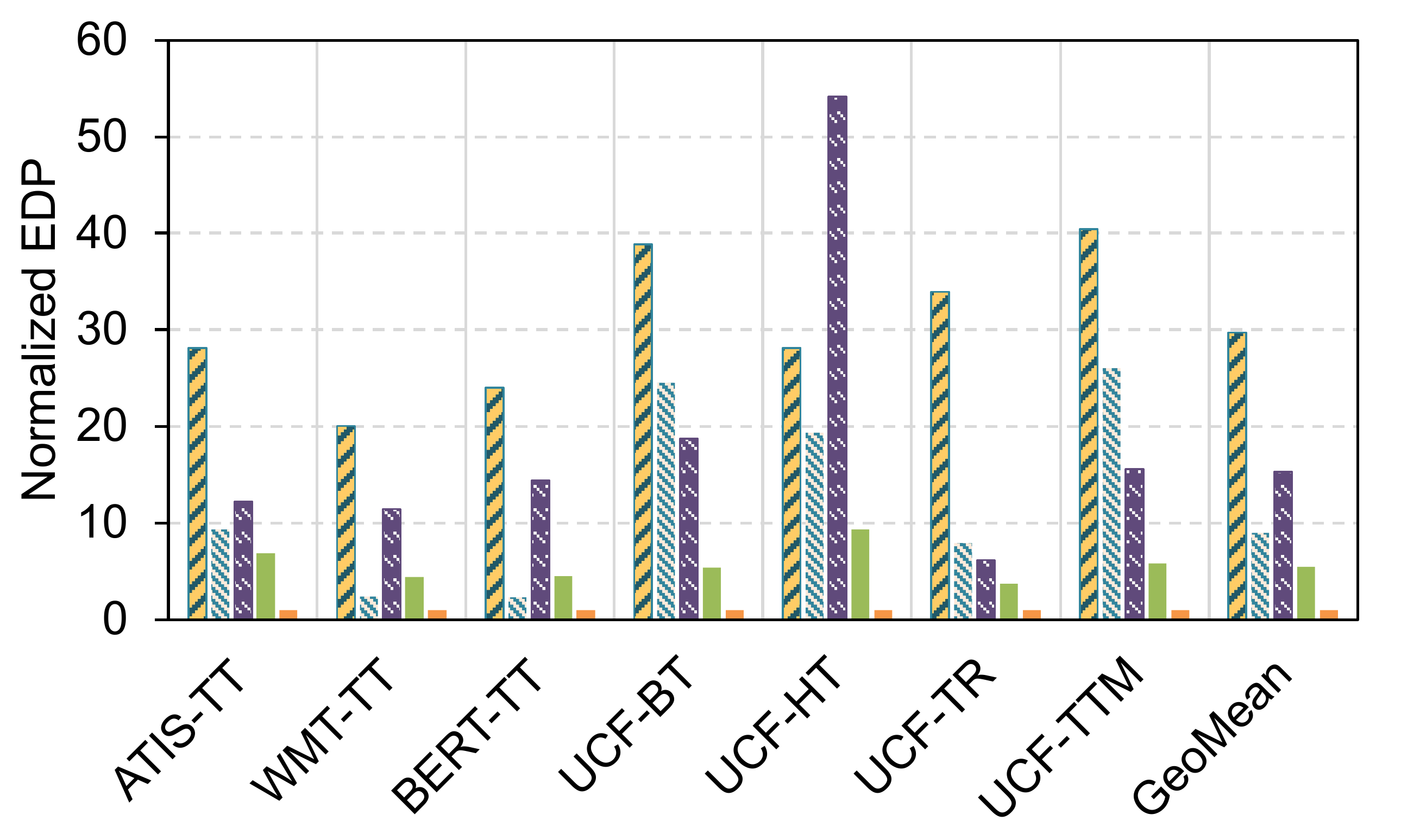}
        \label{subfig:edp_comp}
      }
      \hspace{-10pt}
      \subfigure[]{
      \vspace{-10pt}
      \includegraphics[width=0.24\textwidth]{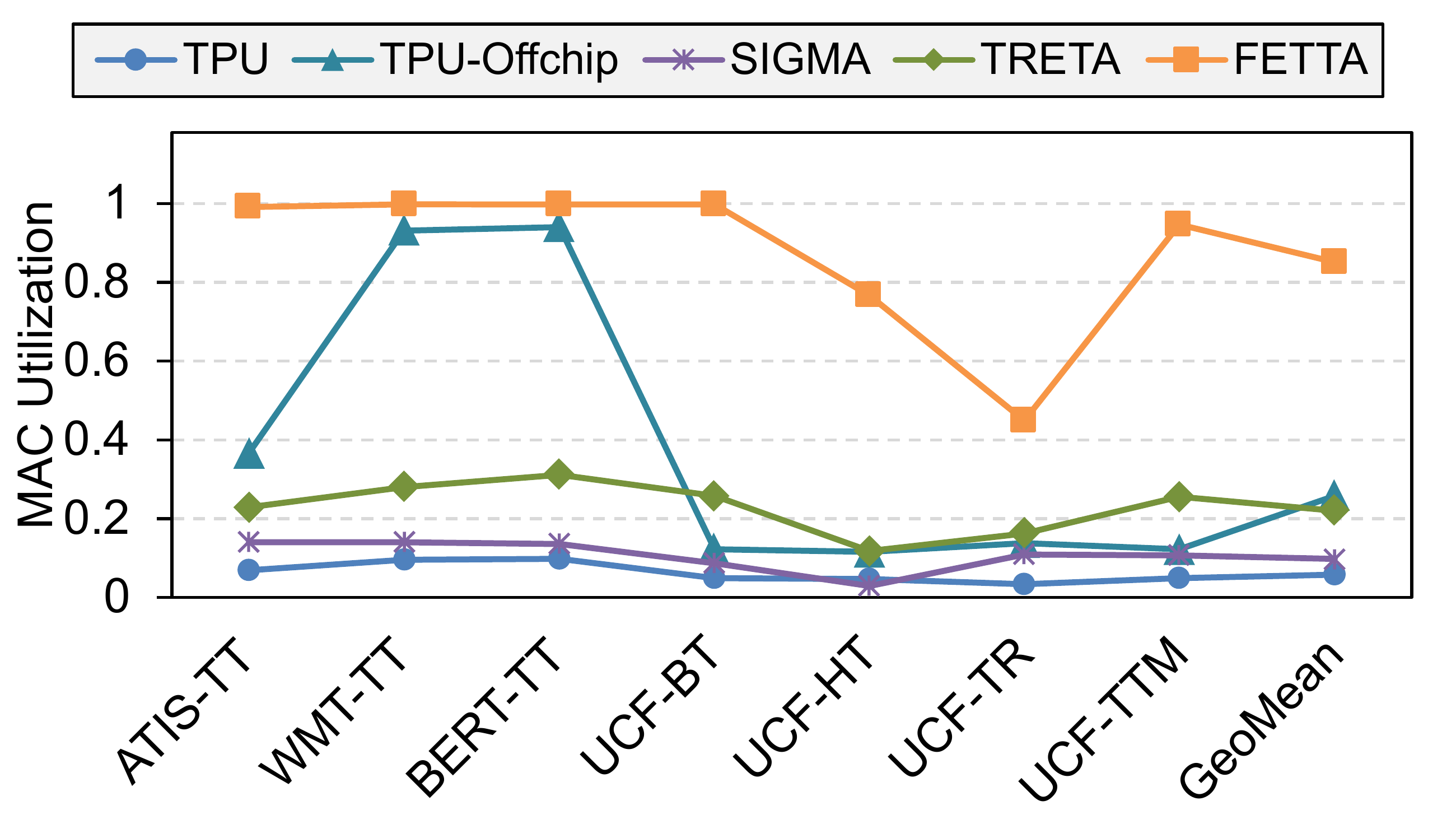}
        \label{subfig:utilization}
      }
      \hspace{-10pt}
      \caption{ Comparison of FETTA and prior training accelerators. (a) Latency, (b) Energy, (c) EDP, (d) MAC utilization}

      \vspace{-10pt}
      \label{fig:res_comp}
    \end{figure}

  {
    Fig.~\ref{fig:dense_comp} illustrates the performance improvement of FETTA over GPU and TPU across various training workloads.
    Both GPU and TPU execute dense and tensorized training with fixed sequences (TNN-F) and CSSE searched sequences (TNN-C), respectively.

    Compared with GPU-Dense, FETTA achieves $1.3\sim144.8\times $ speedup and $69.0\sim1990.9\times$ energy reduction across models with varying compression ratios.
    Under tensorized training, FETTA averagely provides $10.2\times/4.2\times$ speedup and $480.5\times/104.2\times$ energy reduction compared with GPU-TNN-F and GPU-TNN-C, respectively.

    On average, FETTA reduces processing latency by $100.9\times$ / $25.7\times$ / $14.6\times$ and energy consumption by $45.0\times$/$5.6\times$/$2.0\times$ compared with TPU-Dense, TPU-TNN-F, and TPU-TNN-C.
    These gains over TPU-TNN-C primarily result from the flexible architecture and dataflow of FEETA.
    In contrast, TPU employs a weight-stationary loop ordering and fixed parallelism, suffing from an low PE utilization of less than 10\% as shown in Fig.~\ref{subfig:utilization}.
    The performance gains of TPU-TNN-F over TPU-Dense ($3.9\times$ speedup and $8.0\times$ energy reduction) mainly reflect the benefits of model compression.
    The results over TPU-Dense highlight the significant end-to-end performance benefits achieved through algorithm-hardware co-optimization.

    }
    \begin{table}[t]
      \centering
      \caption{ Comparison with Prior Accelerators}
      \label{tab:npu_comp}
      \begin{threeparttable}
        \renewcommand{\arraystretch}{1.2}
        \resizebox{\columnwidth}{!}{
      \begin{tabular}{c|c|c|c|c|c|c}
      \toprule
      \rowcolor{gray!50} \multicolumn{1}{l|}{}      & \bf{FETTA}  & \bf{TPU}    & \bf{SIGMA}  & \bf{TRETA}  & \begin{tabular}[c]{c} \bf{SIGMA} \\ \bf{Sparse} \end{tabular} & \bf{SparSynergy} \\ \hline \hline
      \begin{tabular}[c]{c} Process  [nm] \end{tabular}               & 7      & 7      & 7      & 7      & 28    & 7           \\ \hline
      \rowcolor{row_color_2} \#MAC                      & 256    & 256    & 256    & 256    & 16384 & 512         \\ \hline
      \begin{tabular}[c]{c} Frequency \\ {[}MHz{]} \end{tabular}            & 1000   & 1000   & 1000   & 1000   & 500   & 500         \\ \hline
      \rowcolor{row_color_2} \begin{tabular}[c]{c} Area [mm$^2$]\end{tabular}
                       & 0.21   & 0.19   & 0.24   & 0.25   & \begin{tabular}[c]{c} 65.10 \\ (2.02$^\dagger$) \end{tabular}  & 0.17        \\ \hline
      \begin{tabular}[c]{c} Power [mW] \end{tabular}                 & 175.5 & 140.6 & 336.5 & 312.4 &
                  \begin{tabular}[c]{c} 22330       \\ (8373$^\dagger$)      \end{tabular}  & 353.3      \\ \hline
      \rowcolor{row_color_2} \begin{tabular}[c]{c} Thoughput$^\ddagger$ \\ {[}TOPS{]}  \end{tabular}
                & 3.77   & 0.26   & 0.43   & 0.98   & 10.80 & 0.86        \\ \hline
      \begin{tabular}[c]{c} Power Effi. \\ {[}TOPS/W{]} \end{tabular}  & 21.49  & 1.83   & 1.27   & 3.12   & 1.29$^\dagger$  & 2.44        \\ \hline
      \rowcolor{row_color_2} \begin{tabular}[c]{c} Area Effi. \\ {[}TOPS/mm$^2${]} \end{tabular}
                & 18.02  & 1.33   & 1.79   & 3.93   & 5.35$^\dagger$  & 5.03        \\ \bottomrule
      \end{tabular}
      }
      \begin{tablenotes}
        \footnotesize
        \item[$\dagger$] Normalized to 7 nm technology node based on \cite{Sarangi2021DeepScaleToolAT}.
        \item[$\ddagger$] Average effective throughput over all experiments.
    \end{tablenotes}
    \end{threeparttable}
      \end{table}
    \subsection{Comparison with Prior Accelerators on Tensorized Training}

    To further validate the superiority of our architecture design, we evaluate FETTA against SoTA accelerators on tensorized training workloads shown in Fig.~\ref{fig:res_comp} and Table~\ref{tab:npu_comp}.
    In this configuration, all hardware accelerators perform tensorized training with the CSSE searched optimal contraction sequences.
    Beside, we also compare FETTA with several sparse accelerators, including SIGMA-Sparse \cite{qin2020sigma} and SparSynergy \cite{yang2025sparsynergy}.

    \subsubsection{FETTA vs. TPU-Offchip}

    Since the TPU suffers from a utilization drop caused by data layout inconsistencies, TPU-Offchip mitigates this issue by performing off-chip data layout reordering.
    TPU-Offchip improves the utilization and reduces latency relative to the vanilla TPU implementation.
    However, this improvement comes at the cost of increased energy consumption due to additional DRAM accesses.
    By contrast, FETTA is capable of performing on-chip layout reordering, therefore demonstrating average improvements of $3.30\times$ and $2.73\times$ in speed and energy reduction over TPU-Offchip.

    \subsubsection{FETTA vs. SIGMA}
    SIGMA \cite{qin2020sigma} offers high flexibility, enabling arbitrary spatial mapping shapes. To achieve this, it implements complex distribution networks and provides high on-chip bandwidth for efficient data transportation. However, SIGMA lacks support for data layout reordering, which increases the risk of bank conflicts. As a result, SIGMA exhibits, on average, $8.85\times$ higher latency, $1.73\times$ higher energy consumption, and $15.27\times$ higher energy-delay product compared with FETTA.
    {
      Additionally, FETTA shows smaller area cost due to the simpler on-chip network, leading to an overall $10.1\times$ area efficiency improvement over SIGMA.
    }

    \subsubsection{FETTA vs. TRETA}
    TRETA \cite{shao2023treta} features a hierarchical architecture similar to that of FETTA, theoretically offering sufficient flexibility for dataflow, with a utilization of 22\%.
    However, the absence of flexible distribution and reduction networks necessitates redundant on-chip storage to support its dataflow flexibility, such as data multicasting to multiple CEs.
    Consequently, FETTA achieves $3.86\times$ speedup, $1.41\times$ energy reduction, and $4.58\times$ area efficiency improvement.
    {
    \subsubsection{FETTA vs. Sparse Accelerators}
    Sparse accelerators usually employ complicated encoding, searching, and indexing mechanisms to skip zero values, thereby incurring additional overheads. SIGMA-Sparse supports unstructed sparsity patterns, which brings higher compression ratio as well as requires more complex hardware design. SparSynergy exploits both weight value- and bit-level sparsity.
    Compared with SIGMA-Sparse and SparSynergy, FETTA achieves $16.66/8.81\times$ power efficiency and $3.37/3.58\times$ area efficiency improvement, respectively.
    }

    \begin{figure}[t]
      \centering
      \hspace{-10pt}
      \subfigure[]{
        \includegraphics[width=0.45\linewidth]{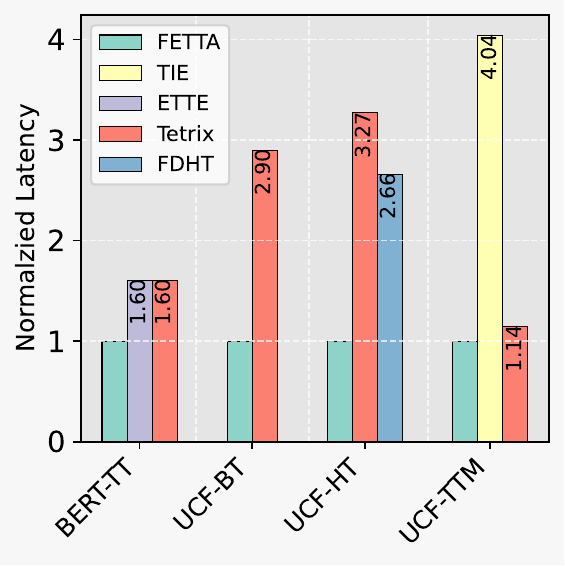}
        \label{subfig:latency_infer}
      }
      \subfigure[]{
        \includegraphics[width=0.45\linewidth]{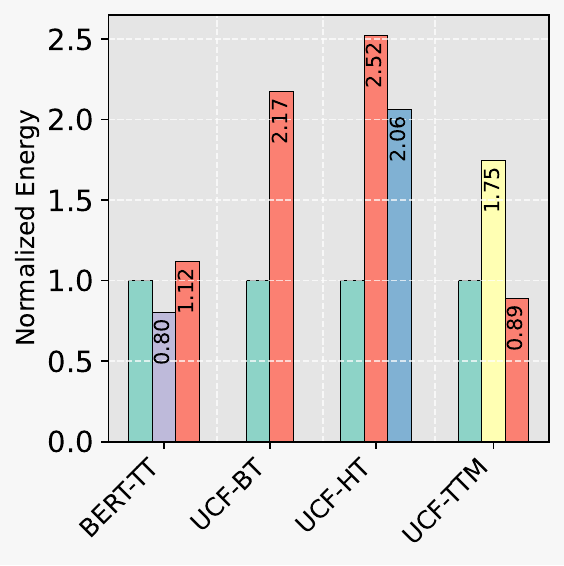}
        \label{subfig:energy_infer}
      }
      \hspace{-10pt}
      \vspace{-5pt}
      \caption{Comparison of FETTA and prior inference accelerators. (a) Latency and (b) Energy.}

      \vspace{-5pt}
      \label{fig:infer_comp}
    \end{figure}

    \subsection{Comparison with TNN Inference Accelerators}
  {
    Fig. \ref{fig:infer_comp} presents a performance comparison between FETTA and prior TNN inference accelerators. Tetrix\cite{zhang2024tetrix} supports contraction paths from its search algorithm and accommodates various tensor formats. In contrast, TIE\cite{deng2019tie} is designed for TTM, ETTE\cite{gong2023ette} is optimized for TT, and FDHT\cite{gong2022algorithm} targets HT. These accelerators execute fixed contraction paths.

    Compared to TIE, FETTA achieves a 4.04$\times$\ speedup while enhancing energy efficiency by 1.75$\times$.
    Against FDHT, it demonstrates a 2.66$\times$\ faster execution with an energy efficiency improvement of 2.06$\times$.
    When evaluated against Tetrix, the performance gain of FETTA varies between 1.14$\times$\ and 3.27$\times$\ in speedup, alongside an energy efficiency improvement ranging from 0.89$\times$\ to 2.52$\times$.

    To support more flexible dataflows and layouts, FETTA introduces additional area and power overheads, leading to slightly higher energy consumption for TTM operations than Tetrix.
    Meanwhile, ETTE, which employs a look-ahead strategy by storing intermediate tensors in registers, benefits from reduced energy consumption.
    Despite this tradeoff, the ability of FETTA to adaptively optimize tensor contractions enables superior performance across various tensor formats.
  }
    \vspace{-5pt}

    \section{Conclusion}\label{sec:conclusion}

    In this paper, FETTA is proposed as a co-design that integrates a flexible hardware architecture with an optimal computing scheme to efficiently perform on-device training of TNNs.
    A CSSE is developed to identify the optimal contraction sequence, maximizing hardware performance and energy efficiency.
    FETTA incorporates a highly flexible and efficient architecture, featuring a reconfigurable CE array designed to support diverse dataflows in TNN training
    Furthermore, transposable butterfly-based distribution and reduction networks are implemented to facilitate flexible tensor shaping operations during computation, achieving seamless dataflow switching while eliminating overhead associated with explicit tensor shaping operations.
    Evaluation results demonstrate that FETTA achieves $20.5\times$/$100.9\times$ speedup and  improves energy efficiency by $567.5\times$/$45.03\times$ energy efficiency compared with GPU and TPU, respectively.
    Furthermore, when compared to prior accelerators for tensorized training workloads, FETTA enhances processing speed by $3.87\sim 14.63\times$, and improves energy efficiency by $1.41 \sim  2.73\times$ on average.

\bibliographystyle{IEEEtran}
\bibliography{ref}

\begin{IEEEbiography}[{\includegraphics[width=1in,height=1.25in,clip,keepaspectratio]{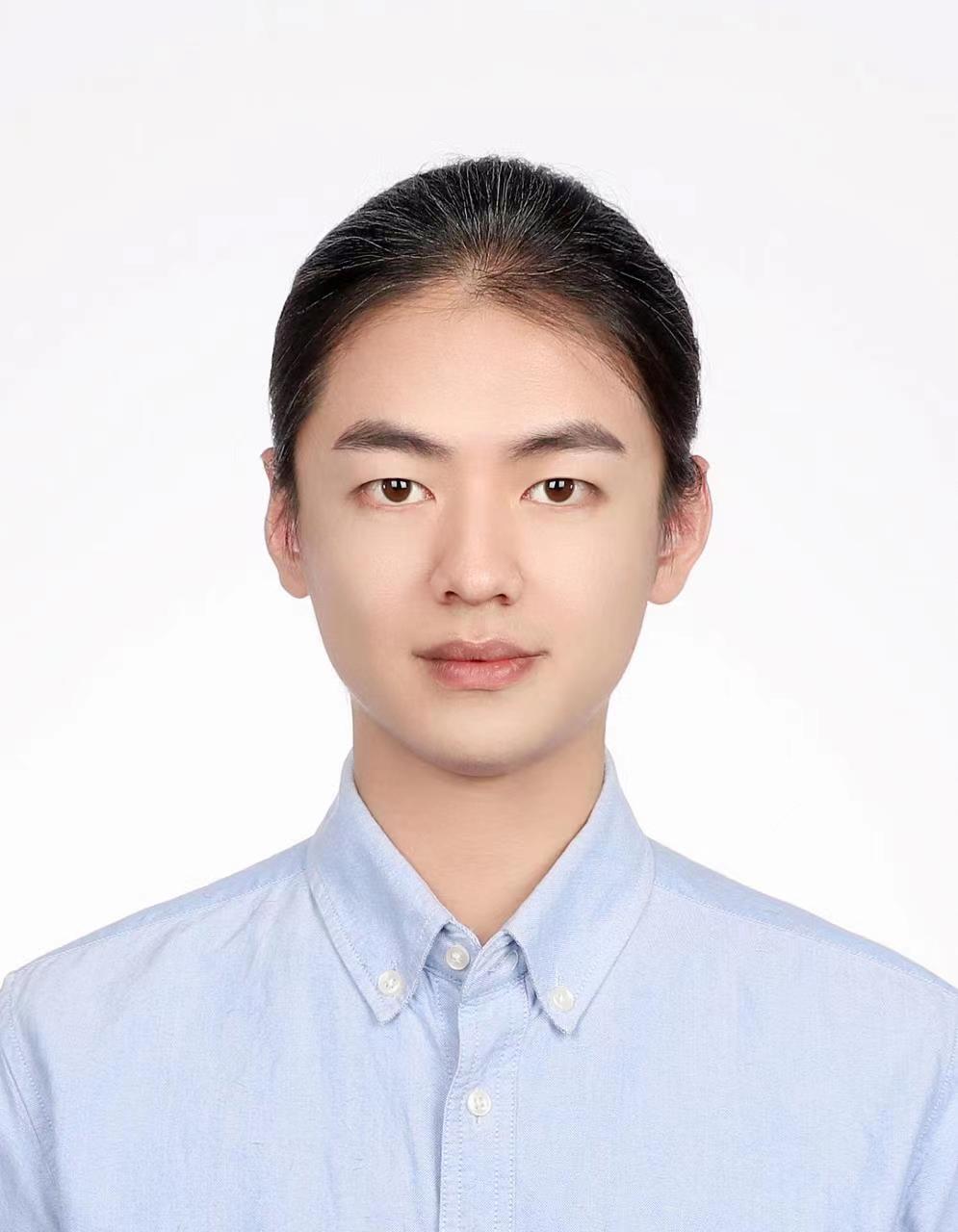}}]{Jinming Lu}
  Jinming Lu received the B.S. degree in microelectronics from Nankai University, Tianjin, China,
  in 2018, and the Ph.D. degree in information and communication engineering from Nanjing University, Nanjing, China, in 2023. Currently, he is a post-doc in University of California, Santa Barbara. His current research interests include automatic speech recognition and deep learning, especially its hardware acceleration and compression algorithms.
  \end{IEEEbiography}

\begin{IEEEbiography}[{\includegraphics[width=1in,height=1.25in,clip,keepaspectratio]{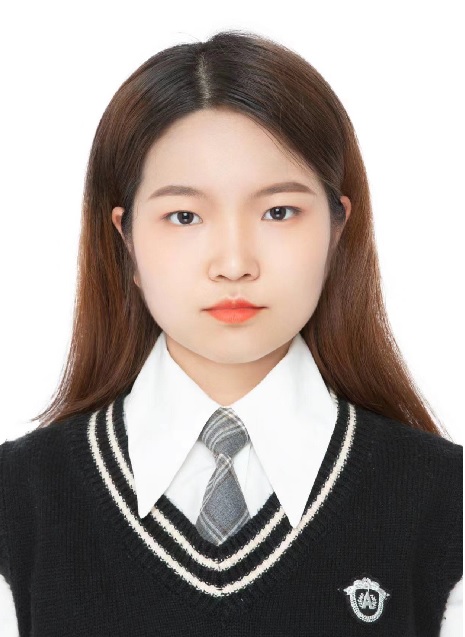}}]{Jiayi Tian}
  Jiayi Tian received the B.Eng. degree in VLSI Design \& System Integration from Nanjing University, Nanjing, China, in 2023, and is currently a Ph.D. student in computer engineering in University of California, Santa Barbara, advised by Prof. Zheng Zhang. Her current research interests are algorithm \& hardware co-design for efficient large language model training and inference.
  \end{IEEEbiography}

  \begin{IEEEbiography}[{\includegraphics[width=1in,height=1.25in,clip,keepaspectratio]{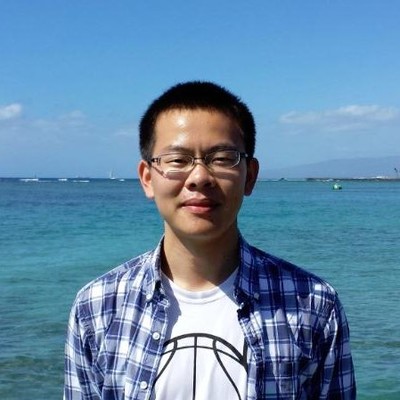}}]{Hai Li}
  Hai Li received the B.S. degree in applied physics from the Huazhong University of Science and Technology, Wuhan, China, in 2011, and the M.S. and Ph.D. degrees in electrical and computer engineering from Carnegie Mellon University, Pittsburgh, PA, USA, in 2015 and 2016, respectively, where he developed foundational theorem of next generation storage system. He joined Intel Corporation in 2016, starting with specialized technologies program. He is currently a Senior Research Scientist with the Exploratory Integrated Circuits, Components Research, Intel Corporation, Hillsboro, OR, USA. He is working on emerging technologies in the joint area of novel device, circuit, and computing paradigm, while managing university collaboration. He serves as Co-Chair of Intel Emerging Technology Strategic Research Sector (SRS).
  \end{IEEEbiography}

  \begin{IEEEbiography}[{\includegraphics[width=1in,height=1.25in,clip,keepaspectratio]{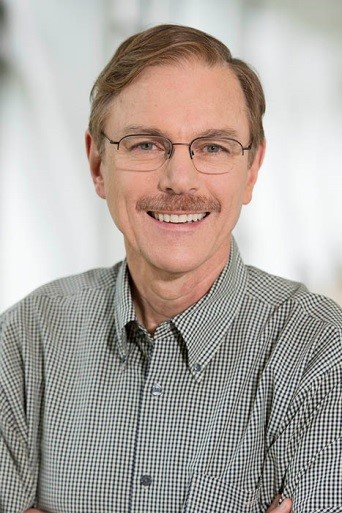}}]{Ian Young}
  Ian. A. Young [M'788, SM96, F'99] is a senior fellow and director of Advanced Circuits and Technology Integration in the Technology and Manufacturing Group at Intel Corporation.
  He received his Bachelor's and Master's degrees in electrical engineering from the University of Melbourne, Australia, in 1972 and 1975. He received his Ph.D. in electrical engineering from the University of California, Berkeley in 1978. He joined Intel in 1983.
  Currently, he is responsible for defining future circuit directions with emerging novel devices and identifying leading options to manufacture energy efficient integrated circuits for computing in the beyond-CMOS era.
  He has authored or coauthored more than 300 technical articles and holds over 250 U.S. patents.
  Dr. Young has received three Intel Achievement Awards. He was a recipient of the 2009 International Solid-State Circuits Conference’s Jack Raper Award for outstanding technology directions paper and the 2018 IEEE Frederik Philips Award “for leadership in research and development on circuits and processes for the evolution of microprocessors.”
  \end{IEEEbiography}

  \begin{IEEEbiography}[{\includegraphics[width=1in,height=1.25in,clip,keepaspectratio]{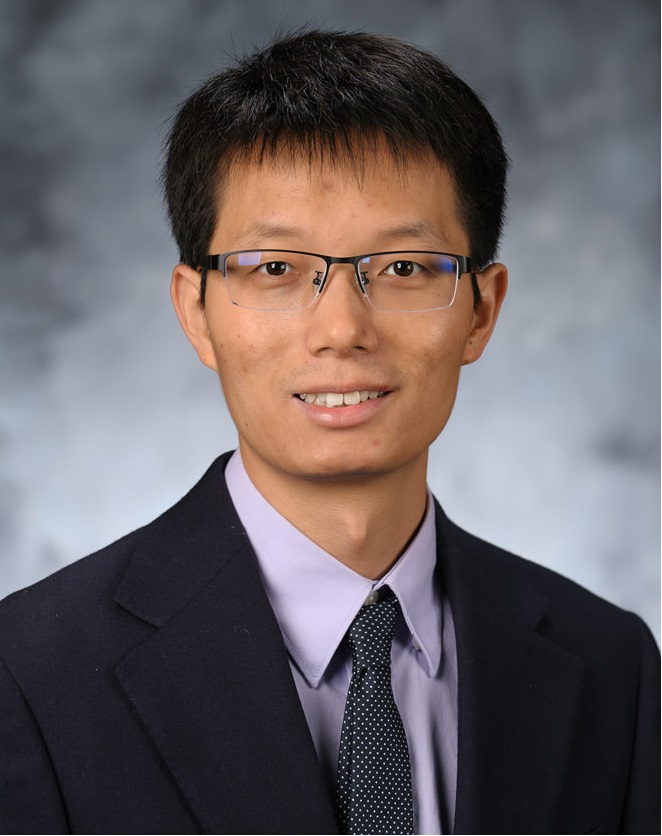}}]{Zheng Zhang}
  Zheng Zhang [M'15] has been an Associate Professor of Electrical and Computer Engineering with the University of California at Santa Barbara (UCSB), since September 2023. He received his Ph.D in Electrical Engineering and Computer Science from the Massachusetts Institute of Technology (MIT), Cambridge, MA, in 2015, M.Phil from the University of Hong Kong in 2010, and B. Eng from Huazhong University of Science and Technology in 2008.
  His research interests include uncertainty-aware design automation for electronics, photonics, and quantum circuits; small-data and data-free scientific machine learning for multi-physics design of 3D IC and chiplet, and tensor-compressed methods for sustainable training of large AI models and for resource-constraint on-device learning.
  \end{IEEEbiography}
\end{document}